\documentclass[fleqn,usenatbib]{mnras}

\usepackage{newtxtext,newtxmath}

\usepackage[T1]{fontenc}
\usepackage{xcolor}
\usepackage{natbib}
\usepackage{lipsum}
\usepackage{caption}
\usepackage{subcaption}

\usepackage{rotating}
\usepackage{longtable}
\usepackage{geometry}
\usepackage{makecell}    
\usepackage{amsmath}     
\usepackage{graphicx}    
\usepackage{booktabs}    
\usepackage{tikz,xcolor,hyperref}

\definecolor{lime}{HTML}{A6CE39}
\DeclareRobustCommand{\orcidicon}{\hspace{-4pt}
	\begin{tikzpicture}
		\draw[lime, fill=lime] (0,0) 
		circle [radius=0.16] 
		node[white] {\hspace{0.1mm}{\fontfamily{qag}\selectfont \tiny ID}};
		\draw[white, fill=white] (-0.07,0.1) 
		circle [radius=0.01];
	\end{tikzpicture}
	\hspace{-3.2mm}
}

\foreach \x in {A, ..., Z}{\expandafter\xdef\csname 
	orcid\x\endcsname{\noexpand\href{https://orcid.org/\csname orcidauthor\x\endcsname}
		{\noexpand\orcidicon}}
}

\usepackage[pagewise]{lineno}
\usepackage{etoolbox}

\DeclareRobustCommand{\VAN}[3]{#2}
\let\VANthebibliography\thebibliography
\def\thebibliography{\DeclareRobustCommand{\VAN}[3]{##3}\VANthebibliography}

\usepackage{graphicx}	
\usepackage{amsmath}	
\usepackage{float}
\def\hb{H$\beta$ }
\def\oiii{[O\,{\sc iii}]~}
\def\oiiia{[O\,{\sc iii}]$\lambda$5007}
\def\oiiias{[O\,{\sc iii}]$\lambda$5007~}
\def\feii{Fe\,{\sc ii}~}
\def\mdot{$\text{M}_\odot$~}
\def\bls{BLSy1s~}
\def\nls{NLSy1s~}
\def\mbh{$\text{M}_{\text{BH}}$~}

\title[]{Estimating the masses of Narrow line Seyfert 1 galaxies using damped random walk method}

\author[Rachana et al.]{
Rachana,$^{1,2}$\orcidA{}\thanks{E-mail: rachana2022@iisc.ac.in}
M. Vivek $^{2}$\orcidB{}\thanks{E-mail: vivek.m@iiap.res.in},
Yue Shen$^{3,4}$\orcidC{}\thanks{shenyue@illinois.edu} 
\\
$^{1}$Joint Astronomy Programme, Department of Physics, Indian Institute of Science, Bangalore 560012, India;\\
$^{2}$Indian Institute of Astrophysics, Block II, Koramangala, Bangalore 560034, India\\
$^{3}$Department of Astronomy, University of Illinois at Urbana-Champaign, Urbana, IL 61801, USA\\
$^{4}$National Center for Supercomputing Applications, University of Illinois at Urbana-Champaign, Urbana, IL 61801, USA}

\date{Accepted XXX. Received YYY; in original form ZZZ}
\pubyear{\the\year{}}

\begin{document}
\label{firstpage}
\pagerange{\pageref{firstpage}--\pageref{lastpage}}
\maketitle

\begin{abstract}
Narrow-line Seyfert 1 galaxies (NLSy1s) are a subclass of active galactic nuclei (AGNs), commonly associated with rapidly accreting, relatively low-mass black holes ($10^6$–$10^8  M_\odot$) hosted in spiral galaxies. Although typically considered to have high Eddington ratios, recent observations, particularly of $\gamma$-ray-emitting NLSy1s, have raised questions about their true black hole masses, with some estimates approaching those of Broad-line Seyfert 1 (BLSy1) systems. In this work, we present the recalibrated mass estimations for a large sample of NLSy1s galaxies with z $<0.8$. We apply the damped random walk (DRW) formalism to a comparison set of 1,141 NLSy1 and 1,143 BLSy1 galaxies, matched in redshift and bolometric luminosity using SDSS DR17 spectroscopy. Our analysis employs a multivariate calibration that incorporates both the Eddington ratio and the rest-frame wavelength to refine the mass estimates. We obtain median DRW-based black hole masses of $\text{log}(M_{\text{BH}}^{\text{DRW}}/M_\odot) = 6.25 \pm 0.65$ for NLSy1s and $7.07 \pm 0.67$ for BLSy1s, in agreement with their respective virial mass distributions. Furthermore, we identify strong inverse trends between the variability amplitude and both optical luminosity and \feii\ emission strength, consistent with a scenario where higher accretion rates suppress long-term optical variability. These findings reinforce the view that NLSy1s harbor smaller black holes and highlight the value of variability-based approaches in tracing AGN accretion properties.
\end{abstract}



\section{Introduction} \label{1}
Narrow-line Seyfert 1 (NLSy1) galaxies represent a unique subclass of active galactic nuclei (AGN) distinguished by their narrow \hb emission lines (FWHM $ \leq 2000$ km s$^{-1}$ ) and relatively weak \oiiias emission lines, with a flux ratio of \oiiia /\hb $\leq$3 \citep{1985ApJ...297..166O,1989ApJ...342..224G}. They typically exhibit steeper soft X-ray spectra (E $<$ 2keV), soft X-ray excess, and rapid soft X-ray variability \citep{1996A&A...305...53B, 1996A&A...309...81W,1999ApJS..125..297L,1999ApJS..125..317L}. These spectral and variability features are found to be significantly correlated. In particular, \citet{1992ApJS...80..109B}, through principal component analysis (PCA), demonstrated that the primary eigenvector (Eigenvector 1) in AGN reflects a strong positive correlation with the strength of \feii emission, an anti-correlation with the strength of \oiii emission and with the width of broad Balmer lines, highlighting the spectral diversity within AGNs. Later, \citet{2014Natur.513..210S} demonstrated the \feii/\hb sequence is indeed correlated with L/$L_\text{Edd}$, making the \feii/\hb ratio a reliable albeit empirical indicator for the Eddington ratio. In addition to this, \citet{2001A&A...372..730V} reported that NLSy1 galaxies frequently exhibit an \feii/\hb intensity ratio exceeding 0.5, reinforcing the role of strong \feii emission as a defining characteristic of this subclass.

In addition to their distinctive spectral and X-ray properties, NLSy1s are also distinguished by the physical characteristics of their central engines. Their relatively narrow widths "broad" emission lines imply lower virial velocities, which, when combined with AGN luminosity, suggest comparatively low black hole masses. These scaling relations, commonly referred to as single-epoch (SE) mass estimates, are established using techniques such as reverberation mapping (RM) \citep{1982ApJ...255..419B, 1993PASP..105..247P, 2021iSci...24j2557C}, which measures the time delay between variability in the continuum and the response in broad emission lines \citep{2000ApJ...533..631K,2005ApJ...629...61K,2006ApJ...651..775B, 2009ApJ...697..160B,2013ApJ...767..149B}. Applying these methods, numerous studies have found that NLSy1s typically host black holes in the range of 10$^6$-10$^8$ $M_{\odot}$, that accrete at high rates, often approaching or exceeding the Eddington limit \citep{2004ApJ...606L..41G,2018MNRAS.480...96W}.

\par
Despite their relatively low black hole masses and high accretion rates, a subset of NLSy1 galaxies has been observed to emit $\gamma$-rays, indicating the presence of relativistic jets \citep{2009ApJ...699..976A, 2011nlsg.confE....F, 2012MNRAS.426..317D, 2018ApJ...853L...2P}. This discovery has sparked significant interest, as it challenges the prevailing notion that powerful jets are exclusive to AGNs with massive black holes in elliptical host galaxies. Detailed multi-wavelength analyses, particularly in the radio band, along with broadband spectral energy distribution modeling, reveal that these $\gamma$-ray emitting NLSy1s share several characteristics with blazars, especially flat-spectrum radio quasars \citep{2016ApJ...819..121P, 2018ApJ...853L...2P}. The key distinction lies in both the host morphology and the inferred black hole mass, as blazars are typically powered by black holes exceeding $10^8$~$M_{\odot}$ and are hosted by elliptical galaxies, whereas NLSy1s are believed to contain lower-mass black holes within spiral galaxies.

\par 
Further evidence challenging virial mass estimates comes from spectro-polarimetric and accretion disk modeling studies. \citet{2016MNRAS.458L..69B} estimated a black hole mass of $6 \times 10^8$~\mdot\ for the $\gamma$-ray-emitting NLSy1 galaxy PKS 2004-447 based on spectro-polarimetric observations, an order of magnitude higher than the $5 \times 10^6$~\mdot\ value obtained from the unpolarised total intensity spectrum. Similarly, \citet{2013MNRAS.431..210C} applied accretion disk model fitting to the spectra of 23 radio-loud NLSy1s (RL-NLSy1s) and found black hole masses comparable to those of blazars. Expanding on this approach, \citet{2019ApJ...881L..24V} analyzed a sample of $\sim$554 RL-NLSy1 and radio-quiet NLSy1 (RQ-NLSy1) galaxies and compared their mass distributions to a control sample of 471 radio-quiet Broad-line Seyfert 1 (RQ-BLSy1) galaxies. They reported that the mean black hole mass for the NLSy1 sample is systematically higher than values derived from virial estimators. Accretion disk modeling, therefore, provides an alternative method for estimating black hole masses in type-1 AGNs \citep{2015MNRAS.446.3427C, 2016MNRAS.460..212C}, while virial estimators remain the standard approach despite their known systematic uncertainties \citep{2018NatAs...2...63M}. However, the caveat here is that most NLSy1s accrete at high accretion rates ($\dot{m} \ge 0.3$, where $\dot{m}=\frac{\dot{M}}{\dot{M}_{Edd}}$), where the standard thin-disk (Shakura–Sunyaev) model is no longer valid, making accretion disk-based mass estimates less reliable in such cases \citep{1988ApJ...332..646A,2003A&A...398..927W}.

\par 
These discrepancies in virial mass estimates highlight the need for alternative approaches and a critical reassessment of existing methods. One such approach is the damped random walk (DRW) model, a stochastic process that effectively describes the optical variability of quasars and AGNs \citep{2009ApJ...698..895K, 2010ApJ...708..927K, 2010ApJ...721.1014M, 2013ApJ...765..106Z}. The DRW model appears to be a good description over months to years timescales, but there are likely deviations from the simple DRW behavior on much shorter or much longer timescales \citep[e.g.,][]{2022MNRAS.514..164S, 2022ApJ...936..132Y}. In practice, the DRW model can be implemented as a Gaussian process with an exponential covariance kernel, enabling efficient modeling of light curve variability through a characteristic damping timescale ($\tau_{\mathrm{DRW}}$) and amplitude ($\sigma_{\mathrm{DRW}}$). In the frequency domain, AGN light curves typically exhibit a power spectral density (PSD) with a slope of -2 at high frequencies and flatten to white noise ($ \text{slope} \rightarrow{0}$) at low frequencies \citep{1999MNRAS.306..637G, 2010ApJ...721.1014M}. The transition between these regimes occurs at a characteristic break frequency, defined as $f_{br} = 1/(2\pi\tau_{\text{DRW}})$, where $\tau_{\text{DRW}}$ denotes the variability damping timescale. This timescale has been proposed as a proxy for physical parameters such as black hole mass. Early studies by \citet{2001ApJ...555..775C} and \citet{2009ApJ...698..895K} suggested a connection between this timescale and the physical properties of the central engine, including black hole mass. However, the exact form of this dependence remains uncertain, with several studies reporting inconsistent or weak correlations \citep[e.g.,][]{2010ApJ...708..927K, 2016A&A...585A.129S}. More recently, \citet{2021Sci...373..789B} reported a statistically robust correlation based on a sample of 67 AGNs with high-quality optical light curves, described by:

\begin{equation}
    \tau_{\text{DRW}} = 107^{+11}_{-12}~\text{days} \left( \frac{M_{\text{BH}}}{10^8\,M_\odot} \right)^{0.38^{+0.05}_{-0.04}}
    \label{burke1}
\end{equation}

where $M_{\text{BH}}$ is the mass of the supermassive black hole (SMBH) and $\tau_{\text{DRW}}$ is the damping timescale characterizing the optical variability. This relation exhibits an intrinsic scatter of $0.33^{+0.11}_{-0.11}$ dex in $M_{\text{BH}}$. Notably, this correlation appears to hold across a wide range of accreting systems. It has been extended to the stellar-mass regime through optical variability studies of nova-like accreting white dwarfs \citep{2015SciA....1E0686S}. Additionally, \citet{2024ApJ...967L..18Z} reported a similar scaling based on a sample of 7 microquasars and 34 blazars observed with the XMM-Newton X-ray Telescope and the Fermi Large Area Telescope, respectively:
\begin{equation}
    \tau = 120^{+15}_{-18}~\text{days} \left( \frac{M_{\text{BH}}}{10^8\,M_\odot} \right)^{0.57 \pm 0.02}
\end{equation}

\par 
Beyond black hole mass, several other physical parameters influence the damping timescale $\tau_{\text{DRW}}$, including the rest-frame wavelength ($\lambda$) and the Eddington ratio ($R_{\text{Edd}}$), defined as the ratio of bolometric luminosity to Eddington luminosity \citep{2010ApJ...721.1014M,2021ApJ...907...96S, 2022MNRAS.514..164S}. The dependence of these parameters can be understood in the context of the standard Shakura–Sunyaev accretion disk model \citep{1973A&A....24..337S}. According to this theory, for a given wavelength $\lambda$, the characteristic emission region $R_\lambda$ satisfies the condition $k_B T(R_\lambda) = hc/\lambda$, where $k_B$ is Boltzmann’s constant, $h$ is Planck’s constant, $c$ is the speed of light, and $T(R_\lambda)$ is the disk temperature at radius ${R}_\lambda$. This leads to the scaling relation:
\begin{equation}
    \frac{R_\lambda}{R_S} \propto M_{\text{BH}}^{-1/3} R_{\text{Edd}}^{1/3} \lambda^{4/3},
\end{equation}
where $R_S = 2GM_{\text{BH}}/c^2$ is the Schwarzschild radius and $R_{\text{Edd}}$ is the Eddington ratio, defined as
\begin{equation}
    R_{\text{Edd}} = \frac{L_{\text{bol}}}{L_{\text{Edd}}} \propto \frac{\dot{M}}{M_{\text{BH}}}
\end{equation}
with $L_{\text{bol}}$ being the bolometric luminosity, $L_{\text{Edd}}$ the Eddington luminosity, and $\dot{M}$ the mass accretion rate.

In standard accretion disk theory, the thermal timescale ($t_{\text{th}}$) is expected to scale with the black hole mass and the disk radius as follows: 
\begin{equation}
t_{\text{th}} \approx 1680 \left(\frac{\alpha}{0.01}\right)^{-1} \left(\frac{M_{\text{BH}}}{10^8 M_\odot}\right) \left(\frac{R}{100 R_s}\right)^{3/2}\text{days},
\end{equation}
where $\alpha$ is the viscosity parameter

In the analysis by \citet{2021Sci...373..789B}, a constant Eddington ratio was assumed across the sample. Under this assumption, the emission radius scales simply with mass as $R_\lambda \propto M_{\text{BH}}^{2/3}$ and hence, the thermal timescale becomes:
\begin{equation}
    t_{\text{th}} \propto M_{\text{BH}}^{1/2},
\end{equation}
implying a direct connection between $\tau_{\text{DRW}}$ and $M_{\text{BH}}$ when Eddington ratio is held fixed. However, for systems with high or variable $R_{\text{Edd}}$, this dependence must be explicitly incorporated.
For AGNs with high or varying Eddington ratios, it is essential to explicitly account for the influence of $R_{\text{Edd}}$ on the thermal timescale. In this regime, the scaling becomes:
\begin{equation}
    t_{\text{th}} \propto \alpha^{-1} \lambda^2 M_{\text{BH}}^{1/2} R_{\text{Edd}}^{1/2}.
\end{equation}
Building on this theoretical framework, \citet{2024ApJ...966....8Z} empirically calibrated the dependence of $\tau_{\text{DRW}}$ on $M_{\text{BH}}$, $R_{\text{Edd}}$, and rest-frame wavelength ($\lambda_{\text{rest}}$) using a sample of 190 quasars. They derived the following relation based on DRW modeling:
\begin{align}
    \log_{10}(\tau / \text{days}) = & \, a \log_{10}(M_{\text{BH}} / M_{\odot}) + b \log_{10}(R_{\text{Edd}}) \notag \\
    & + c \log_{10}(\lambda_{\text{rest}} / \text{\AA}) + d,
    \label{zhou}
\end{align}
where the best-fitting parameters are $a = 0.65 \pm 0.01$, $b = 0.65 \pm 0.01$, $c = 1.19 \pm 0.01$, and $d = -6.04 \pm 0.05$ (all at $1\sigma$ uncertainty). This multi-parameter calibration demonstrates that $\tau_{\text{DRW}}$ is sensitive not only to black hole mass, but also to accretion rate and emission wavelength, especially in high-Eddington ratio systems. In this formulation, the accretion ratio is defined as
\[
R_{Edd} = \frac{L_{\text{bol}}}{(1+k)L_{\text{Edd}}}, \quad k = \tfrac{1}{3},
\]
where the factor (1+k) accounts for additional energy dissipation associated with magnetic activity in the corona relative to turbulence within the accretion disk \citep{2020ApJ...891..178S,2024ApJ...966....8Z}. Motivated by this, we revisit the black hole mass estimates for NLSy1 galaxies, which are known to accrete at high Eddington ratios, often near or above the Eddington limit. Using the DRW framework, we aim to recalibrate $M_{\text{BH}}$ for a large sample of NLSy1s by explicitly incorporating the role of accretion rate and comparing the results with their BLSy1 counterparts. Our analysis is based on $g$-band photometric light curves from the Zwicky Transient Facility (ZTF), including both archival and forced photometry, with sources drawn from the catalog of \citet{2024MNRAS.527.7055P}. Section~\ref{2} details the sample selection and light curve assembly. In Section~\ref{3}, we describe the DRW modeling approach using the \texttt{taufit} package and the construction of the final analysis sample. Section~\ref{4} presents our main results, including updated mass distributions and variability correlations, followed by conclusions in Section~\ref{5}.

\section{Sample Selection}\label{2}

We began with an initial sample of 22,656 NLSy1s and 52,273 BLSy1s from the catalog compiled by \citet{2024MNRAS.527.7055P}, based on spectroscopic data from the Sloan Digital Sky Survey (SDSS) Data Release 17. These sources, classified as either QSOs or GALAXIES by the SDSS pipeline, span redshifts up to $z < 0.8$. To ensure reliable NLSy1 classification, the catalog required that both the \hb and \oiii emission lines be present in the optical spectra, with the FWHM of the broad \hb line less than 2000 km/s and a flux ratio \oiii/\hb\ $\leq$ 3 \citep{1985ApJ...297..166O, 1989ApJ...342..224G}.

\begin{figure}
    \centering
    \includegraphics[width=1\linewidth]{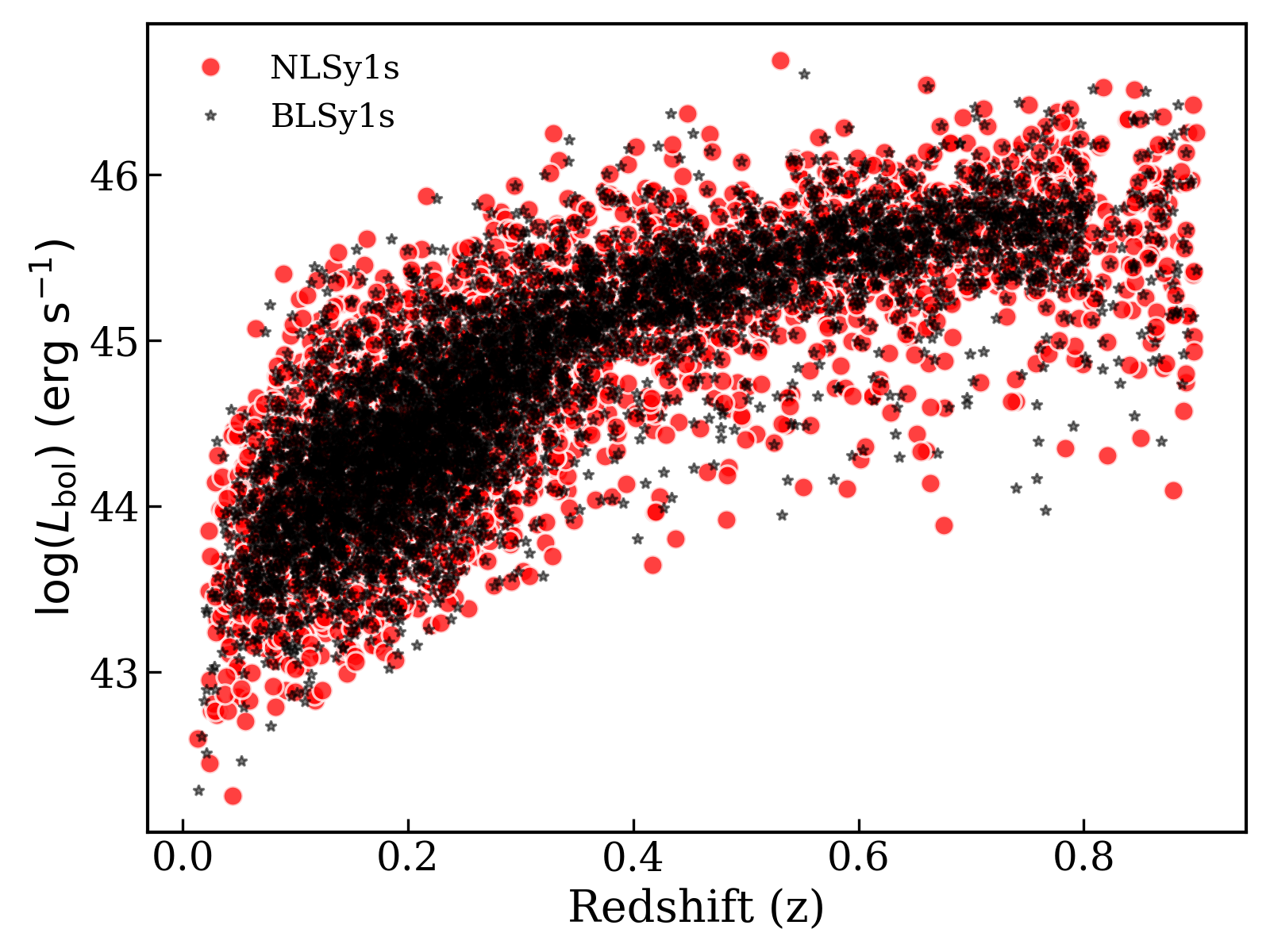}
    \caption{Distribution of the refined sample of NLSy1s (Red) and their matched BLSy1 counterparts (Black) in the luminosity–redshift (L–$z$) plane. All sources have SDSS $r$-band magnitudes $\leq 19$ and were selected from the catalog of \citet{2024MNRAS.527.7055P}. The BLSy1 sample was matched using a maximum distance of 0.2 in the L–$z$ plane.}
    \label{fig: scatter lz}
\end{figure}

For this work, we selected a refined sample of 5,666 NLSy1s and a controlled sample of 5,557 BLSy1s, matched in the Redshift-Luminosity (L–$z$) plane and limited to sources brighter than 19 magnitudes in the SDSS $r$ band (see Fig.~\ref{fig: scatter lz}). The BLSy1 counterparts were selected such that the matching distance in the L–$z$ plane did not exceed 0.2, resulting in a well-controlled sample with over 98$\%$ of NLSy1s successfully matched. The SDSS $r$-band filter was chosen for the magnitude cut because it has a relatively high throughput, compared to other SDSS filters. To analyze the optical variability of these sources, we utilized light curves from both the archival ZTF survey and the ZTF forced photometry dataset \citep{2023arXiv230516279M}. The forced photometry data provide a significantly greater number of epochs and a longer observational baseline than the archival data, thereby improving the robustness and reliability of the DRW model fitting. 

\subsection{Photometric Data}
We retrieved the light curves for our final sample from the Infrared Science Archive (IRSA) at IPAC, using the most recent public data release (DR23) of the ZTF survey \citep{2019PASP..131g8001G,2019PASP..131a8002B,2019PASP..131a8003M}. The data were obtained using a positional search radius of 1 arcsec, which also includes intranight observations from overlapping field IDs. ZTF is a new optical time-domain survey conducted primarily with the 48-inch Samuel Oschin Schmidt telescope at the Palomar Observatory. It employs a custom-built wide-field camera with a field of view of 47 deg$^2$ and observes in the g, r, and i bands, with effective wavelengths of 4753, 6369, and 7915~\text{\AA}, respectively. In this study, we focus exclusively on the $g$-band light curves, which provide the highest number of observations and the longest temporal baseline ($\sim$2000 days). This makes it particularly well-suited for studying AGN optical variability, compared to the more sparsely sampled r and i-band data. Also, we do not use the r and i band light curves to explore the wavelength dependence on $\tau$, as the small separation between the $g$ and $r$ bands limits the ability to identify any significant trends in $\tau$ with rest-frame wavelength. To ensure robust variability modeling, we refined the initial sample by selecting only those sources with $g$-band light curves containing more than 150 epochs. This selection yielded a subsample of 4968 NLSy1 and 5,044 BLSy1 galaxies, which were subjected to further DRW fitting and quality assessment.

\begin{figure*}
    \centering
    \vspace*{2em}
    \includegraphics[width=0.48\linewidth]{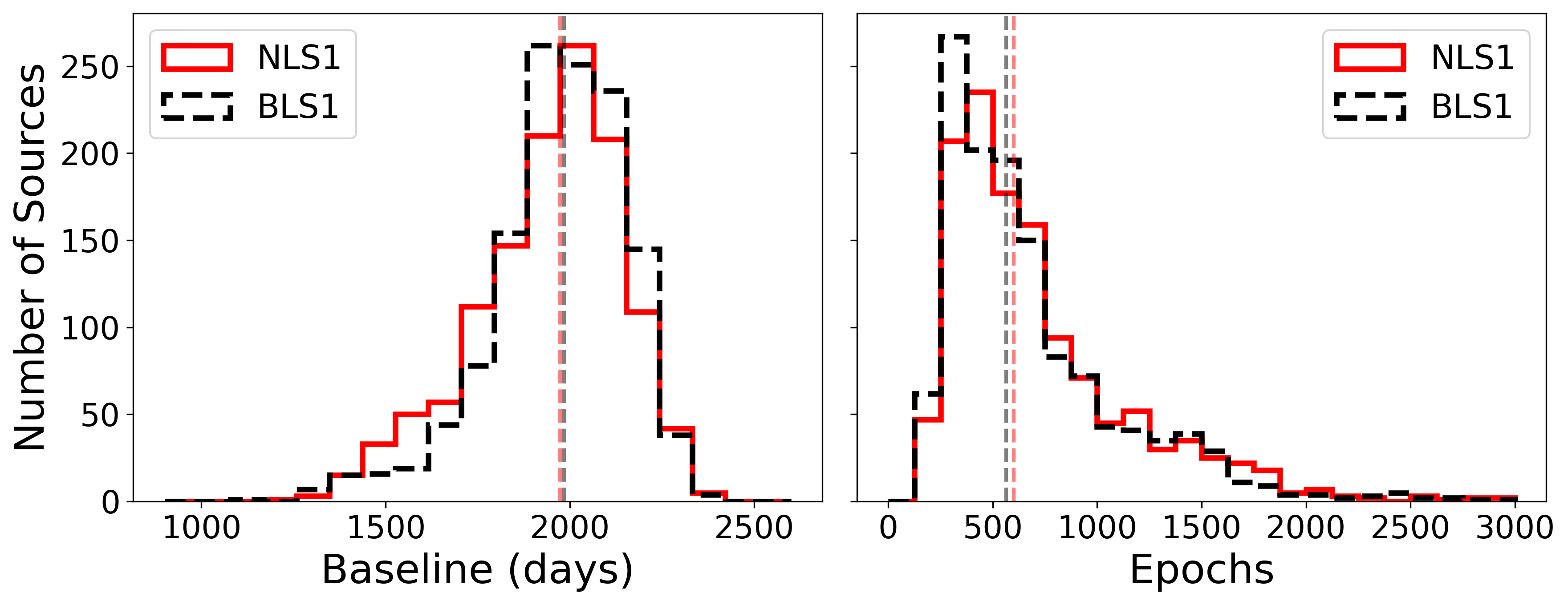}
    \hfill
    \includegraphics[width=0.48\linewidth]{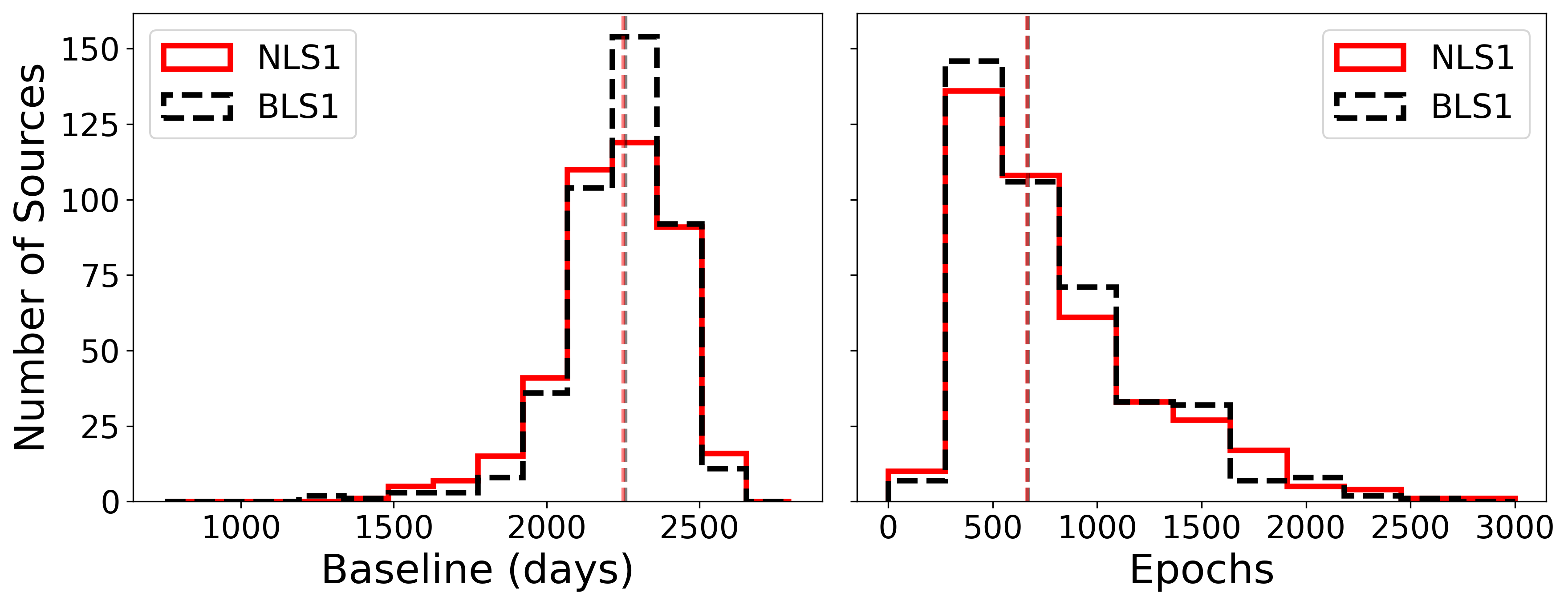}
    \caption{Distribution of the number of epochs and temporal baselines for the ZTF $g$-band light curves. The left panel shows the archival (normal) photometry, while the right panel shows the forced photometry. Red solid-line histograms represent NLSy1 galaxies (final sample: 1,254), and black dashed-line histograms represent BLSy1 galaxies (final sample: 1,270). For the archival light curves, the median baseline is $\sim$2,000 days with a median of 580 high-quality epochs. In comparison, the forced photometry offers an extended baseline of $\sim$2,250 days and a higher median of $\sim$665 epochs, enabling improved sampling for variability studies.}
    \label{fig:epoch_baseline_normal}
\end{figure*}

\par 
Before applying the DRW modeling to the archival light curve data, we performed 3$\sigma$ clipping on each seasonal segment of the light curves. This procedure helped mitigate the impact of outliers by removing extreme deviations within individual seasons, thereby preserving the intrinsic variability of the sources. In addition, we filtered the light curves by retaining only those epochs with \textit{CATFLAGS} = 0, which ensures the exclusion of data points affected by known observational issues such as moonlight contamination, poor image subtraction, saturation, or proximity to charged-coupled device edges. This quality filtering step enhances the reliability of the photometric measurements used in our analysis. The archival ZTF photometric light curves typically contain around $\sim$580 high-quality epochs spanning a rest-frame baseline of approximately $\sim$2000 days, with a typical photometric uncertainty of 0.047 magnitudes in an image. The distribution of epochs and baseline lengths is illustrated in Fig.~\ref{fig:epoch_baseline_normal}.

\subsection{Forced Photometry Data}
We also obtained data from the Zwicky Transient Facility Forced Photometry Service (ZFPS), maintained by the ZTF Science Data System team at IPAC/Caltech \citep{2023arXiv230516279M}, which provides light curves on a request basis. The ZTF forced photometric light curves typically consist of an average of $\sim$665 epochs spanning a baseline of approximately $\sim$2250 days, with a typical photometric uncertainty of 0.035 magnitudes per detection. In forced photometry, the flux of a source is measured at fixed celestial coordinates across all available epochs, regardless of whether the source was independently detected in individual exposures. The forced photometry is carried out on difference images, obtained by subtracting the reference frame from each science exposure, rather than on the original science images. The measured differential flux is then combined with the reference flux to obtain the total calibrated flux. This method ensures consistent flux estimation across time and enables the recovery of faint or transient features. Since the ZFPS includes both recent and archival observations, it is particularly well-suited for generating uniform light curves of variable AGN and improving the reliability of DRW modeling.

\par
Following ZFPS guidelines\footnote{\url{https://irsa.ipac.caltech.edu/data/ZTF/docs/ztf_forced_photometry.pdf}}, we performed a series of quality control steps on all forced photometry light curves. Key photometric quantities, such as difference flux and reference magnitude, were converted to flux space, with associated uncertainties propagated accordingly. We computed total fluxes and retained only those epochs with a signal-to-noise ratio (SNR) greater than 3 to ensure the reliability of the measurements. Calibrated magnitudes were then derived, and non-detections or unreliable points were discarded. We restricted the analysis to $g$-band observations and applied a filter on the \textit{PROCSTATUS} flag, retaining only values of 0, 59, or 60, which indicate successful or minimally problematic reductions. The \textit{PROCSTATUS} flag value of 0 indicates successful execution, while 59 and 60 correspond to minor processing flags, slightly elevated photometric uncertainties. These flags do not significantly compromise the quality of the light curves and were found to preserve the reliability of variability measurements without introducing systematic bias. To further mitigate contamination from poor image subtractions or extended structures, we excluded epochs with anomalously high values of \textit{NEARESTREFCHI}, which quantifies the reduced chi-squared of the PSF fit to the nearest reference image; higher values typically indicate poor fits due to nearby sources or extended morphology. As with the archival light curves, we also applied 3$\sigma$ clipping on a seasonal basis to remove statistical outliers, ensuring consistent treatment across both datasets. These steps yielded clean, high-quality light curves suitable for DRW modeling and time-domain variability analysis.

\begin{figure*}
    \centering
    \includegraphics[width=1\linewidth]{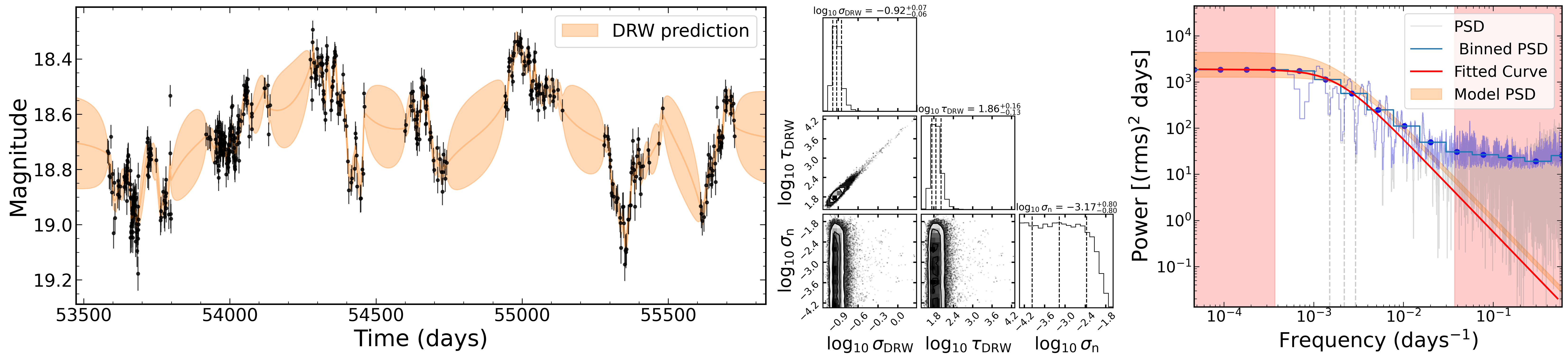}
    \caption{\textbf{DRW modeling of NLS1 SDSS J000144.85+110959.9} \textit{Left panel:} The g-band light curve of SDSS J000144.85+110959.9 (\(M_{\rm BH} = 10^{7.703}\,M_{\odot}\)) DRW model is fitted with 1\(\sigma\) uncertainty as orange shaded area. \textit{Right panel}: Power spectral density (PSD) normalized and binned, shown with $1\sigma$ errors. The best-fit broken power law is overplotted in red and the DRW-based posterior prediction of the PSD is represented by the orange shaded band. The corresponding break frequency \(f_{br}\) (from the broken power law fit) and \(1/(2 \pi \tau_{DRW})\) (from the DRW fitting). The red shaded regions correspond to timescales greater than 20$\%$ the light curve length (in panels left lower and right lower) and less than the mean cadence (panel C), where the PSD is not well sampled.}
    \label{fig:psd}
\end{figure*}

\begin{figure}
    \centering
    \includegraphics[width=1\linewidth]{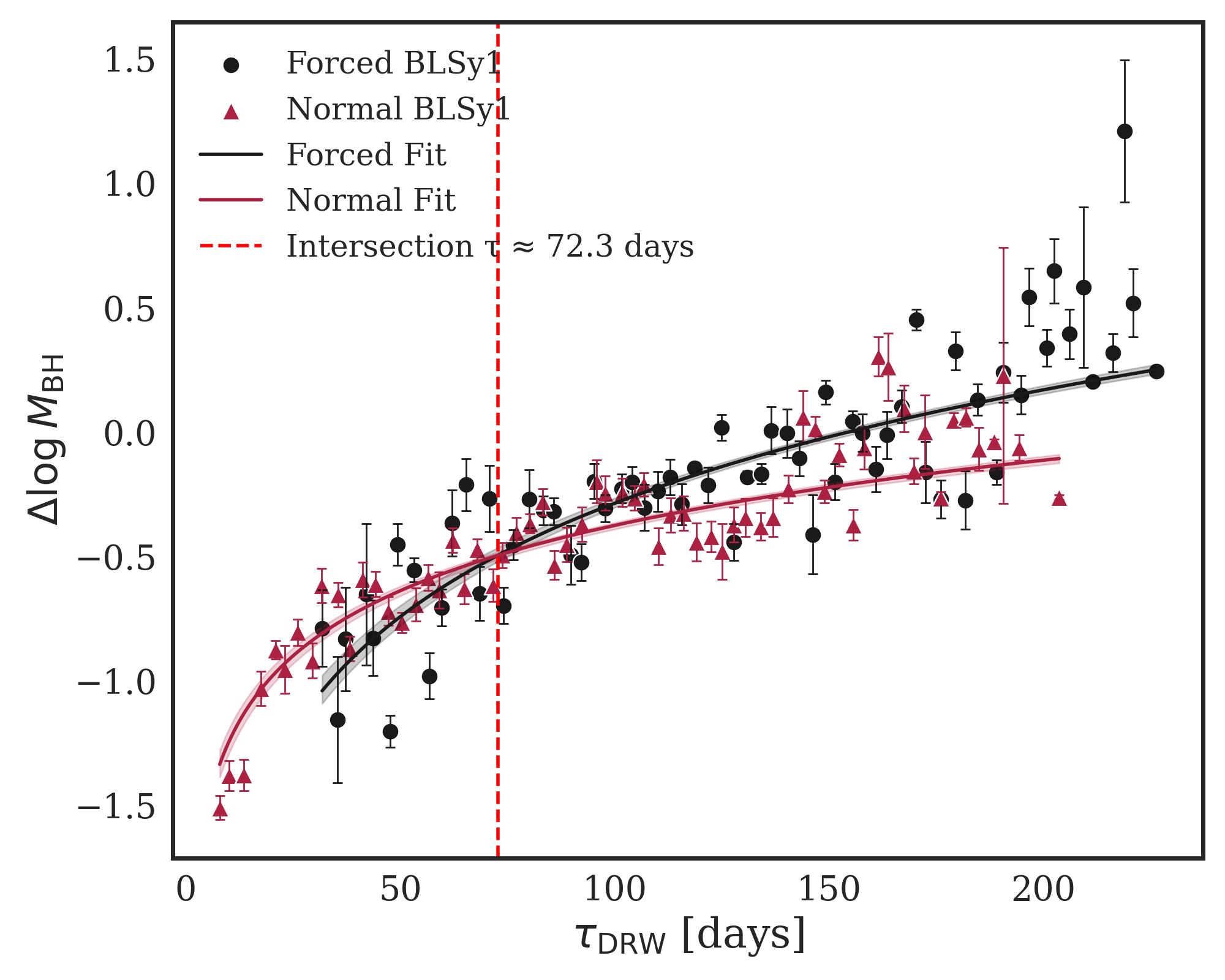}
    \caption{Comparison of the black hole mass offset, defined as $\Delta \log M_{\mathrm{BH}} = \log M_{\mathrm{BH, Zhou}}^{\text{DRW}}$(equation~\eqref{zhou}) - $\log M_{\mathrm{BH}}^{\text{SE}}$, as a function of the DRW timescale $\tau_{\mathrm{DRW}}$, for forced and normal BLSy1 light curves. The associated error bars correspond to the median absolute deviation scaled by $\sqrt{N}$ in each bin. The solid black and red lines show the best-fit MCMC regression models for the forced and normal samples, respectively, with shaded regions indicating the $\pm3\sigma$ confidence intervals. The vertical red dashed line marks the intersection point of the two fits at $\tau_{\mathrm{DRW}} \approx 72.3$ days, suggesting a potential threshold in timescale-dependent mass discrepancies between the two sampling regimes.}
    \label{fig:limit_70}
\end{figure}

\section{Data Analysis: Light Curve Fitting}\label{3}
AGN light curves have been extensively studied using the DRW framework, which characterizes variability through two key parameters: the damping timescale, $\tau$ representing the signal decorrelation timescale, and the variability amplitude, $\sigma$ \citep{2010ApJ...721.1014M}. We modeled the light curves using the DRW-based \texttt{taufit} package\footnote{\url{https://github.com/burke86/taufit}}, developed by \citet{2021Sci...373..789B}. The package is built on the fast Gaussian process solver \texttt{Celerite} \citep{2017AJ....154..220F}, which uses the following covariance kernel:

\begin{equation}
    k(t_{ij})=2\sigma_{\text{DRW}}^2 \text{exp}(-t_{ij}/\tau_{\text{DRW}}) + \sigma_{i}^2\delta_{ij}
    \label{kernel}
\end{equation}
where $t_{ij} = |t_i - t_j|$ is the rest-frame time interval between epochs $t_i$ and $t_j$. The term where $\sigma_i$ captures additional white noise that is common to all epochs, often referred to as jitter, to compensate for underestimated or unmodeled uncertainties in the photometric data, with $\sigma_i^2$ representing its variance. The Kronecker delta $\delta_{ij}$ ensures that the noise is uncorrelated between different epochs. This covariance structure corresponds to the form of the structure function (SF), given by:

\begin{equation}
    \text{SF}^2_{ij}=SF^2_{\infty} (1- \text{exp}(-t_{ij}/\tau_{\text{DRW}}))
    \label{SF}
\end{equation}

where $SF_{\infty} = \sqrt{2}\sigma_{\text{DRW}}$ denotes the asymptotic variability amplitude. In the frequency domain, this kernel gives rise to a PSD that transitions from a flat (white noise) regime at low frequencies to a power-law ($f^{-2}$) decline at high frequencies. The characteristic break frequency at which this transition occurs is defined as:

\begin{equation}
f_{\text{br}} = \frac{1}{2\pi \tau_{\text{DRW}}},
\end{equation}
which provides a direct link between the time-domain damping timescale and the frequency-domain behavior of the variability.

Within \texttt{taufit}, we also perform Markov chain Monte Carlo (MCMC) sampling via the \texttt{EMCEE} package \citep{2013PASP..125..306F}, to explore the posterior distributions of the model parameters. We adopted uniform priors for all parameters, used the median of each posterior distribution as the best-fit value, and estimated the $1\sigma$ uncertainties from the 16th and 84th percentiles. In our analysis, these priors are uniform (flat) in log-space for the main parameters: the damping timescale ($\tau_{\rm DRW}$), variability amplitude ($\sigma_{\rm DRW}$), and white-noise term ($\sigma_{\rm i}$). The prior ranges are automatically set by \texttt{taufit} based on the light-curve duration, cadence, and photometric precision:

\begin{align}
\log a &\in [\log(0.001\, \text{(mag err)}_{\rm min}),\, \log(10\,A)], \nonumber\\ 
\log c &\in [\log(1/(10\,\text{baseline})),\, \log(1/\text{cadence}_{\rm min})], \nonumber\\
\hspace{0.32cm}\log \sigma_{\rm i} &\in [-10,\, \log A] \nonumber
\end{align}

where $\text{mag err}_{\rm min}$ is the minimum photometric error, $A$ is the total amplitude of variability in the light curve, defined as 
$A = (\text{mag} + \text{mag err})_{\rm max} - (\text{mag} - \text{mag err})_{\rm min}$, \textit{baseline} is the total observing period, and $\textit{cadence}{_{\rm min}}$ is the minimum sampling interval. These parameters further constrain the physical quantities of interest, where $c$ corresponds to the damping timescale (c =$1/\tau_{\rm DRW}$) and $a$ corresponds to the variability amplitude term ($a = 2\sigma_{DRW}^2$).

\par
In order to reliably constrain the damping time scale ($\tau_{\text{DRW}}$) while fitting the light curves, it is important that the duration of the light curve is significantly longer than $\tau_{\text{DRW}}$ as pointed out by \citep{2017A&A...597A.128K}. For light curves shorter than a few times $\tau_{\text{DRW}}$, the measured $\tau_{\text{DRW}}$ can be systematically biased low and saturated around 20–40\% of the light curve duration, with elevated scatter in the measurements. Furthermore, the amplitude of variability $\sigma$ should also exceed the photometric measurement uncertainties, as emphasized by \citet{2021Sci...373..789B}. This ensures that for low SNR regimes, the inferred $\tau_{\text{DRW}}$ does not exceed its true intrinsic/true value due to noise-driven fluctuations.

\subsection{Final sample}
We fitted and analyzed the light curves for each source in our sample using the DRW model described above. Fig.~\ref{fig:psd} illustrates the results for the NLS1 galaxy SDSS J000144.85+110959.9. The left panel shows the observed light curve along with the predicted curve based on the best-fitting parameters. The middle panel displays a corner plot of the DRW parameters, $\sigma_{\text{DRW}}$, $\tau_{\text{DRW}}$, and $\sigma_{i}$, obtained using \texttt{taufit}. The right panel presents the PSD of the observed light curve in gray, with the model PSD shown as an orange shaded region representing the 1$\sigma$ posterior uncertainty. Dashed vertical lines mark the break frequency and its corresponding 1$\sigma$ confidence interval.


We applied the following criteria to select our final sample:
\begin{enumerate}
    \item Following \citet{2017A&A...597A.128K}, we required that each light curve span at least ten times the damping timescale, i.e., baseline $>$ $10 \times \tau_{\text{DRW}}$, to ensure robust temporal coverage and minimize biases. 
    
    \item The best-fitting $\tau_{\text{DRW}}$ must exceed the median cadence to prevent undersampling related bias \citep{2021Sci...373..789B}. Additionally, we ensured that the upper bound (84th percentile) of the posterior distribution for $\tau_{\text{DRW}}$ remained within the light curve duration.
    
    \item We retained only those sources where the intrinsic variability amplitude ($\sigma_{\text{DRW}}$) exceeded the total noise. Specifically, we required $\sigma_{\text{DRW}}^2 > \sigma_i^2 + dy^2$, where $\sigma_i$ is the jitter component and $dy$ is the median photometric uncertainty of the light curve.

    \item Each light curve must include data from at least four distinct ZTF observing seasons, with each season comprising a minimum of 50 epochs. On average, light curves in the final sample span seven seasons and cover a $\sim$2000-day baseline.
    
    \item We excluded sources whose auto-correlation functions were statistically consistent with Gaussian white noise within the predicted $3\sigma$ confidence band (p-value $<$ 0.05).
    
\end{enumerate}

\begin{figure*}
    \centering
    \begin{minipage}[b]{0.44\textwidth}
        \centering
        \includegraphics[width=\linewidth]{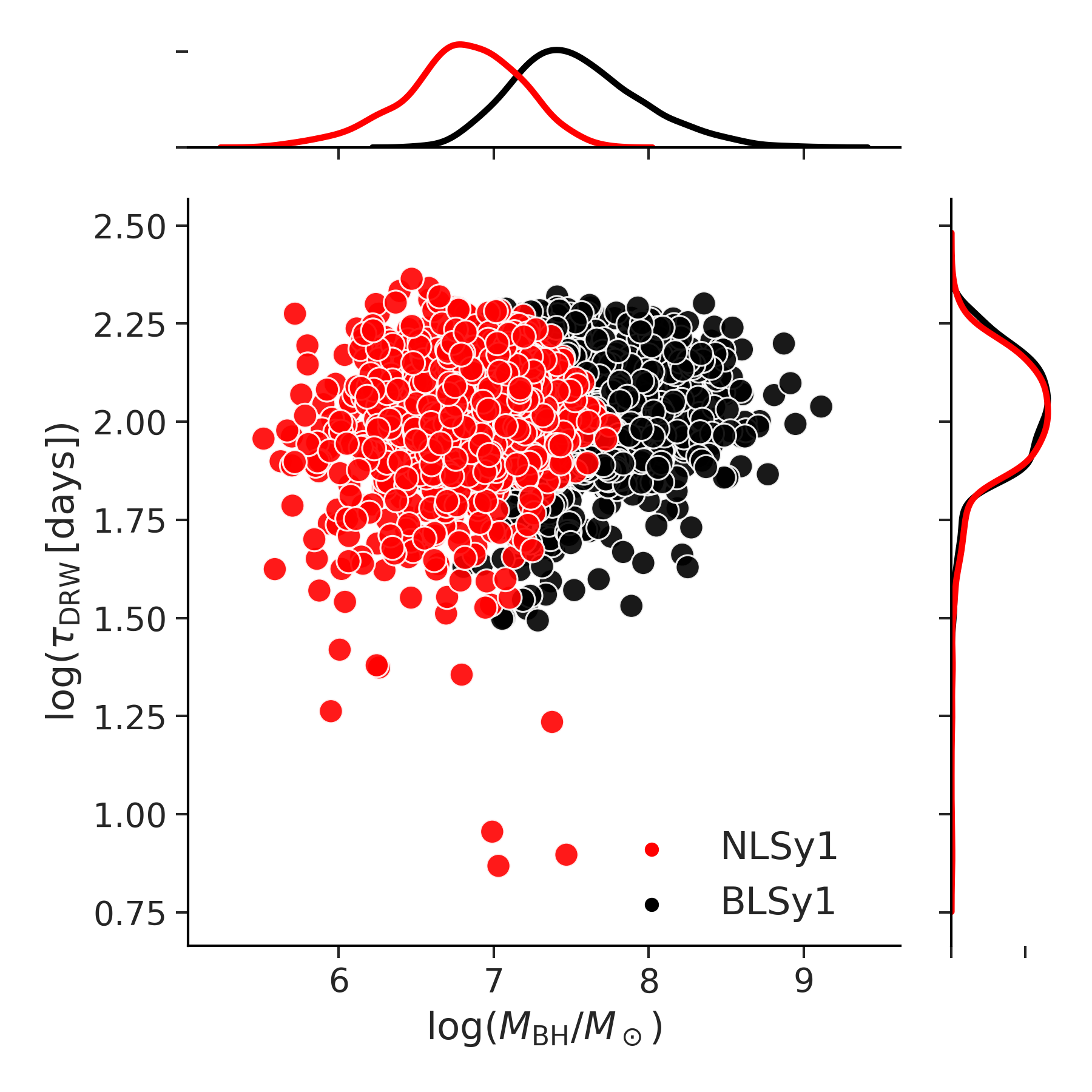}
        \caption{ The DRW damping timescale ($\tau_{\text{DRW}}$) as a function of single-epoch black hole mass estimates ($M_{\text{BH}}$) for the final sample of \nls (red) and \bls (black) galaxies. \textit{Right}: The corresponding distributions of $\tau_{\text{DRW}}$, with solid red and dashed black lines for NLSy1s and BLSy1s, respectively. \textit{Upper}: The corresponding distributions are single epoch mass $M_{\text{BH}}$, for NLSy1s (red) and BLSy1s (black).}
        \label{fig: scatter colin}
    \end{minipage}%
    \hspace{1.5cm}
    \begin{minipage}[b]{0.44\textwidth}
        \centering
        \includegraphics[width=\linewidth]{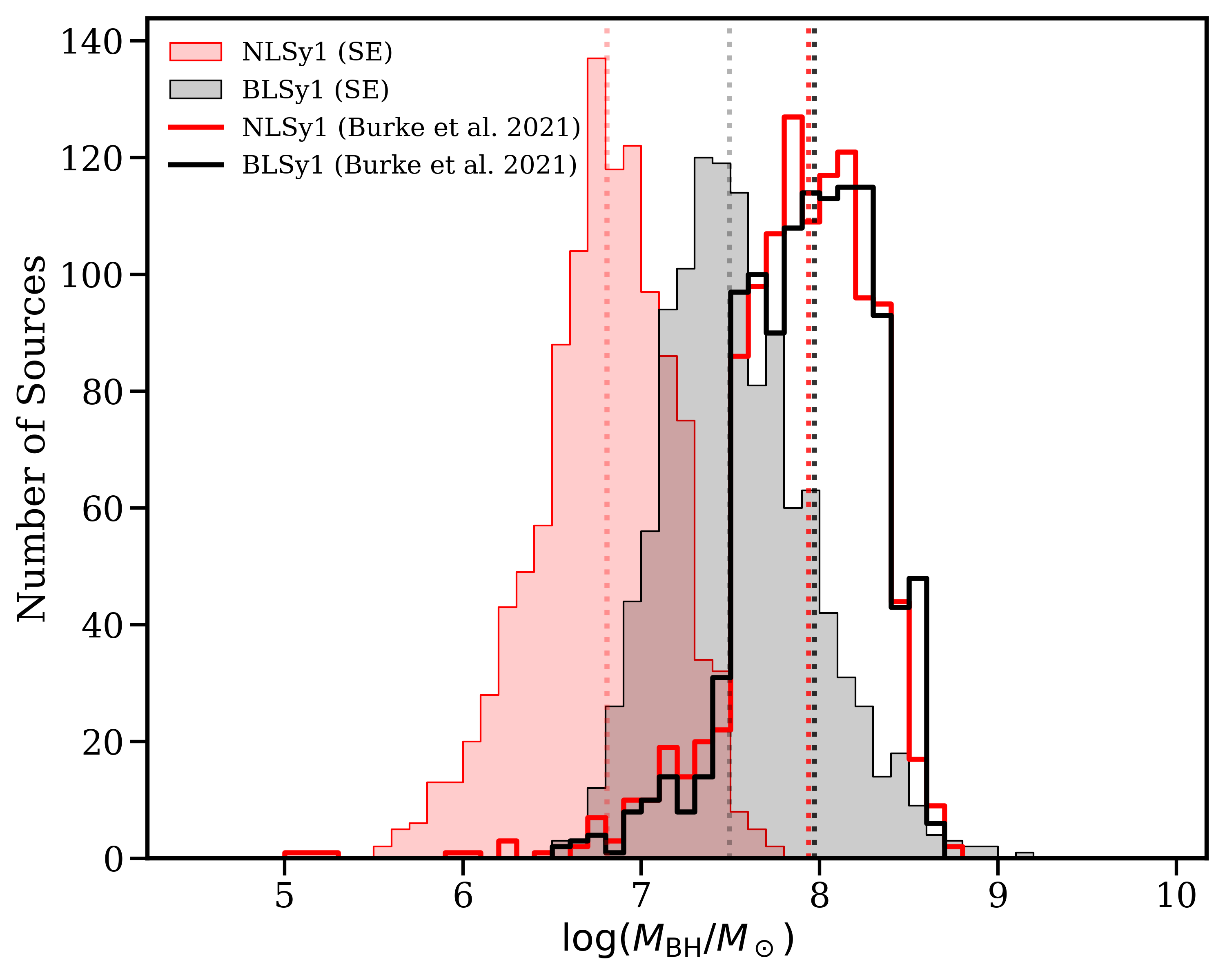}
        \vspace{0.5cm} 
        \caption{Distribution of black hole masses (log \mbh) in units of $M_\odot$ for the final sample of NLSy1 (red) and BLSy1 (grey) sources. The single-epoch mass estimates for NLSy1s (shaded red) and BLSy1s (shaded grey) are shown. Solid lines represent mass estimates from the DRW model (as described in equation~\ref{burke1}), for NLSy1s (red) and BLSy1s (black). The dashed vertical lines indicate the median values.}
        \label{fig: colin burke distribution}
    \end{minipage}
\end{figure*}

After applying all selection criteria and estimating black hole masses from the DRW parameters using equation~\ref{burke1}, we obtain a distribution of $\log_{10}(M_{\mathrm{BH}}/M_\odot)$ that shows a distinct long tail toward lower masses. This tail consists almost entirely of sources with very short damping timescales, $\tau_{\text{DRW}} \lesssim 70$ days. For these cases, the PSD derived from the DRW model appeared nearly flat, as illustrated in Fig.~\ref{fig:photometry_comparison_all}. This is most likely due to shorter baseline coverage and sparse sampling in the archival light curves. Such conditions yield poorly constrained DRW parameters, often producing artificially flat model predictions that fail to capture the intrinsic variability. In addition, for the BLSy1 sub-sample, even after incorporating Eddington ratios when calculating black hole masses using the equation~\ref{zhou} relation, the mass offset ($\Delta \log M_{\mathrm{BH}} = \log M_{\mathrm{BH, Zhou}}^{\text{DRW}}$(equation~\eqref{zhou}) - $\log M_{\mathrm{BH}}^{\text{SE}}$) shows significant deviations from the expectations based on single-epoch virial estimates (Fig.~\ref{fig:limit_70}), again indicating that these DRW fits are biased by insufficient data quality.

To address this issue, we retrieved updated light curves from the ZTF Forced Photometry Service (ZFPS) for approximately 307 sources with $\tau_{\text{DRW}} \leq70$ days, incorporating all available observations up to 16 June 2025. Forced photometry offers denser temporal coverage and extended baselines, including an additional ZTF observing season. Although forced photometry may underestimate photometric uncertainties, particularly for highly variable or slightly extended sources, its improved temporal sampling enhances the reliability of DRW modeling. The revised light curves were processed following the procedures described in \citet{2019PASP..131a8003M} and refitted using the same DRW formalism applied to the archival data. A comparison of DRW-based black hole masses from forced and archival photometry is presented in Fig.~\ref{fig:limit_70}. As seen in Fig.\ref{fig:photometry_comparison_all}, for these same sources the updated fits yield improved damping timescales, and the PSDs are no longer flat, indicating that the intrinsic variability is now better recovered. Hence, for sources with $\tau_{\text{DRW}} \leq 70$ days, forced photometry yields systematically more consistent and physically plausible $\Delta \log M_{\rm BH}$ estimates. However, for longer timescales, discrepancies between the two datasets begin to emerge, likely due to growing biases in forced photometric uncertainties.

Additional AGN physical parameters, such as black hole mass, redshift, \feii strength, and luminosity, are adopted from \citet{2024MNRAS.527.7055P}. After applying all selection criteria, we obtained a final sample of 1,141 NLSy1s and 1,143 BLSy1s. The two samples remain well matched in the luminosity–redshift ($L$–$z$) plane. We find that 81\% of these final NLSy1 sources have corresponding BLSy1 counterparts within a matching distance of 0.2. For each source, we compiled key parameters, including the light curve baseline, median cadence, damping timescale ($\tau_{\text{obs}}$), variability amplitude ($\sigma_{\text{DRW}}$), photometric noise ($\sigma_i$), bolometric luminosity ($L_{\mathrm{bol}}$), black hole mass ($M_{\mathrm{BH}}$), Eddington ratio ($\lambda$), and redshift ($z$). A representative table listing these parameters for a subset of these sources is provided in Appendix~\ref{tab:table}.

\section{Results}\label{4}
We applied the DRW modeling framework to the $g$-band light curves of 1,141 NLSy1 and 1,143 BLSy1 galaxies using both archival and forced photometry data from the ZTF survey. For each source, we obtained posterior estimates of the DRW parameters, $\tau_{\mathrm{DRW}}$, $\sigma_{\mathrm{DRW}}$, and $\sigma_i$. Among these, $\tau_{\mathrm{DRW}}$ was subsequently used for black hole mass estimation.
%

\subsection{The dependence of $\tau_{\mathrm{DRW}}$ on $M_{\text{BH}}$}
To investigate the relationship between DRW damping timescales and black hole mass, we first compared $\tau_{\mathrm{DRW}}$ for NLSy1 and BLSy1 galaxies with their single-epoch virial mass estimates. Fig.~\ref{fig: scatter colin} presents this comparison, where red points correspond to NLSy1 galaxies and black points to BLSy1 galaxies. The top and right panels show the corresponding histograms of $\log M_{\mathrm{BH}}^{\mathrm{SE}}$ and $\tau_{\mathrm{DRW}}$, respectively. Although the two samples exhibit nearly identical $\tau_{\mathrm{DRW}}$ distributions, their $\log M_{\mathrm{BH}}^{\mathrm{SE}}$ distributions are clearly distinct, with NLSy1s peaking at lower masses. We then estimated DRW-inferred black hole masses by applying the empirical $\tau_{\mathrm{DRW}}$–$M_{\mathrm{BH}}$ relation calibrated by \citet{2021Sci...373..789B}, given in equation~\ref{burke1}. Fig.~\ref{fig: colin burke distribution} shows the distributions of DRW-inferred black hole masses for NLSy1 and BLSy1 galaxies. The solid red and black lines represent the NLSy1 and BLSy1 populations, respectively. The median values are $\log(M_{\mathrm{BH}}^{\mathrm{DRW}}/M_{\odot}) = 7.94 \pm 0.38$ for NLSy1s and $7.97 \pm 0.36$ for BLSy1s. For comparison, the median single-epoch virial mass estimates are $6.81 \pm 0.38$ for NLSy1s and $7.49 \pm 0.41$ for BLSy1s, as shown by the red and black shaded histograms in Fig.~\ref{fig: colin burke distribution}, respectively.

\par
The DRW-based black hole mass distributions for NLSy1s and BLSy1s appear similar, but are systematically higher in both cases than the single-epoch virial estimates. As discussed in Section~\ref{1}, DRW damping timescales depend on both black hole mass and Eddington ratio. However, the calibration by \citet{2021Sci...373..789B} assumes a constant Eddington ratio, and neglecting this dependence can lead to overestimated black hole masses in high-$R_{\mathrm{Edd}}$ systems.

\subsection{The dependence of $\tau_{DRW}$ on $M_{\text{BH}}$ and $R_{\text{Edd}}$}
In our sample, $R_{\text{Edd}}$ spans a wide range, from 0.03 to 6.54 for NLSy1s and from 0.001 to 0.95 for BLSy1s, underscoring the need to account for accretion-rate dependence when modeling DRW timescales. To address this, we adopt the multivariate calibration from \citet{2024ApJ...966....8Z} (equation~\ref{zhou}), which explicitly incorporates both $M_{\text{BH}}$ and $R_{\text{Edd}}$ into the DRW framework. We also account for the wavelength dependence of variability timescales by using the $g$-band central wavelength (4753~\AA), corrected to the rest-frame value as $\lambda_{\mathrm{rest}} = 4753 \text{\AA}/(1 + z)$.
\par 
$R_{\mathrm{Edd}}$ is typically determined from the bolometric luminosity and the black hole mass. Since our goal is to examine black hole masses in NLSy1s, using the catalog $R_{\mathrm{Edd}}$ introduces a cyclic bias, as these values are derived from single-epoch virial masses that may be underestimated due to orientation effects. Consequently, the inferred $R_{\mathrm{Edd}}$ values for NLSy1s may also be biased. In BLSy1s, where $R_{\mathrm{Edd}}$ spans a narrower range, the mass–damping timescale correlation in equation~\ref{burke1} remains broadly applicable. Several studies have shown that the strength of the optical \feii emission, quantified by R4570, serves as a more inclination-independent proxy for the Eddington ratio (e.g., \citealt{2002ApJ...565...78B}; \citealt{2014Natur.513..210S}). Here, R4570 is defined as the ratio of the \feii flux in the wavelength range 4434–4684~\AA\ to the flux of the broad component of the \hb line. In our analysis, we use R4570 as a proxy for $R_{\mathrm{Edd}}$ and explore three strategies to establish a parametric relation between the two.

\begin{figure*}
    \centering
    \begin{minipage}[b]{0.47\textwidth}
        \centering
        \includegraphics[width=\linewidth, height=6.5cm, keepaspectratio]{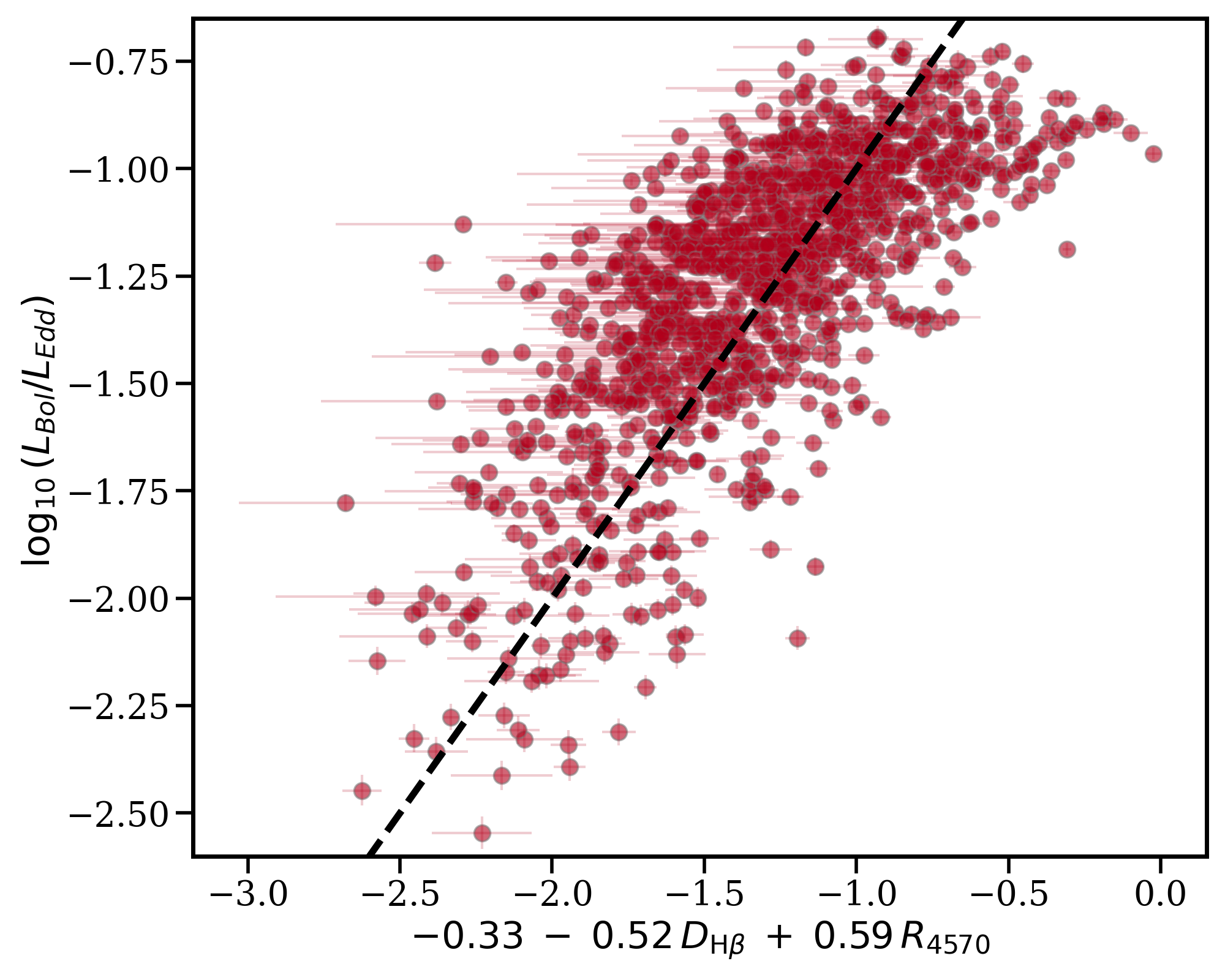}
        \caption{Observed versus modeled Eddington ratios for the BLSy1 sample. The modeled values of $\log_{10}(L_{\mathrm{Bol}}/L_{\mathrm{Edd}})$ are derived from the best-fitting linear combination of the H$\beta$ line shape parameter ($D_{\mathrm{H\beta}}$) and R4570. Each point corresponds to an individual source, with horizontal error bars showing measurement uncertainties. The black dashed line marks the one-to-one relation for reference.}
        \label{fig:du_recalibrated}
    \end{minipage}%
    \hspace{0.2cm}
    \begin{minipage}[b]{0.47\textwidth}
        \centering
        \includegraphics[width=\linewidth, height=6.5cm, keepaspectratio]{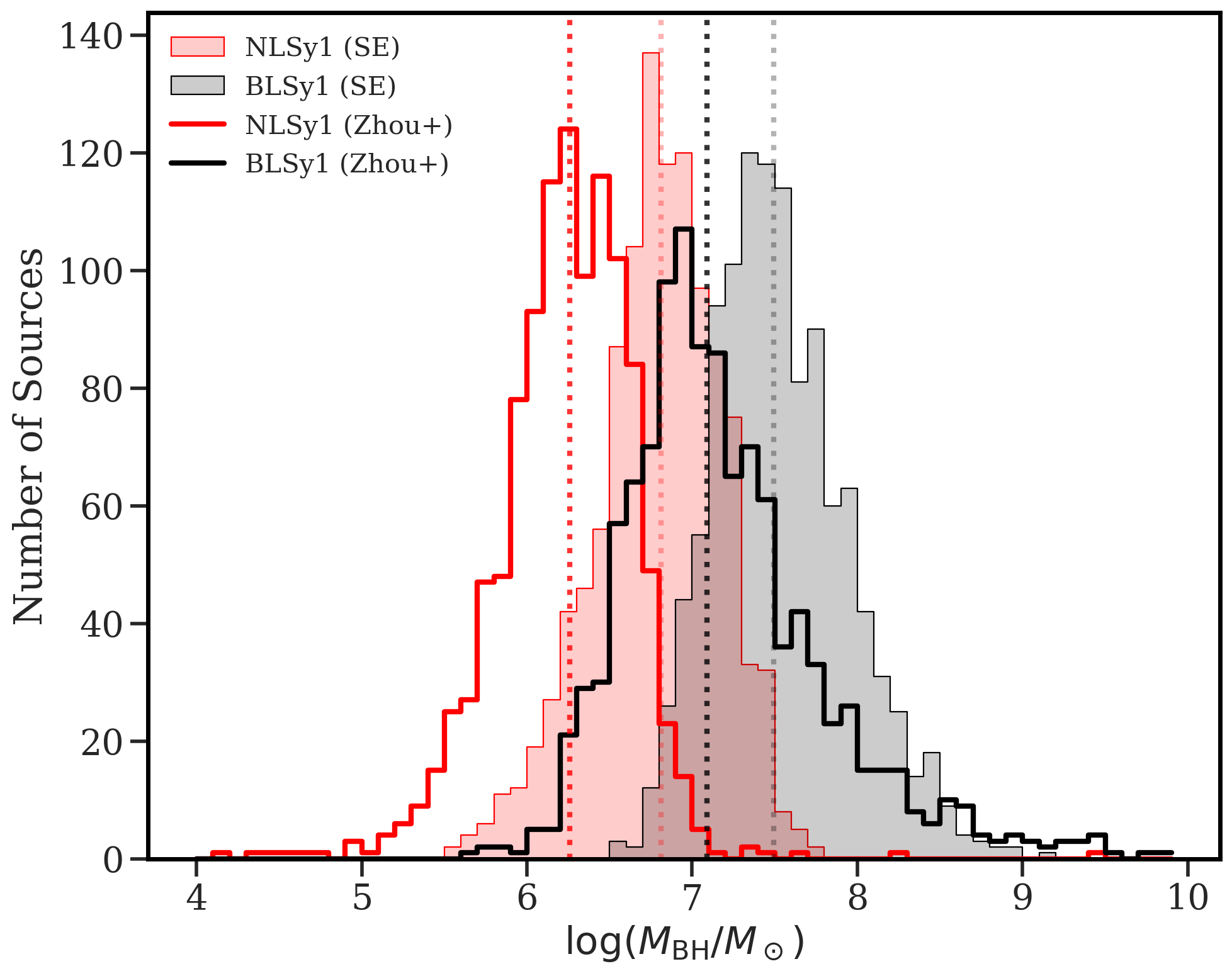}
        \vspace{0.5cm} 
        \caption{Distribution of black hole masses ($\log M_{\rm BH}$) in units of $M_\odot$ for the final sample of NLSy1 (red) and BLSy1 (black) sources. The single-epoch mass estimates (shaded) are shown along with DRW-based estimates (solid lines). The dashed vertical lines indicate the median values.}
        \label{fig:zhou_distribution}
    \end{minipage}
\end{figure*}

First, we fit a linear relation between R4570 and $R_{\mathrm{Edd}}$ for the BLSy1 sample, for which the catalog $R_{\mathrm{Edd}}$ values should be relatively free of any biases. Fig.~\ref{fig:FeII_strength} shows the scatter plot of $\log R_{\mathrm{Edd}}$ versus R4570 for the full sample of BLSy1 (red) and NLSy1 (blue) galaxies. For the BLSy1s, R4570 was binned in steps of 0.05, and a linear fit to the binned data yields, $\log_{10}$(R$_{\mathrm{Edd}})$ = 1.66 $\times$ \, R4570 - 1.372. The best-fit line for the BLSy1s is shown as a solid black line, with its extrapolation to higher R4570 indicated by the dashed black line.
While the fit broadly traces both populations, the substantial scatter, even within the BLSy1 sample, demonstrates that a single-parameter relation between R4570 and 
$R_{\mathrm{Edd}}$ cannot capture the underlying dependencies. Comparing the \citet{2024MNRAS.527.7055P} catalog $R_{\mathrm{Edd}}$
for the BLSy1 sample with the values reconstructed from the linear relation, reveals a pronounced discrepancy between their distributions. Consequently, we refrain from using this relation to estimate black hole masses. Nevertheless, our analysis emphasizes that the $R_{\mathrm{Edd}}$-R4570 connection is inherently multivariate.

Next, we used the multivariate correlation established by \citet{2016ApJ...818L..14D} between $\log R_{\mathrm{Edd}}$, R4570, and $D_{\mathrm{H}\beta}$, where $D_{\mathrm{H}\beta}$ is defined as the ratio of the FWHM to the line dispersion of the \hb\ profile. This approach extends beyond the simple linear R4570–$R_{\mathrm{Edd}}$ relation by also accounting for the \hb\ line shape, thereby reducing scatter and potential biases in the estimated Eddington ratios. The relation is expressed as

\begin{equation}
\log_{10}(R_{\mathrm{Edd}}) = 0.31 - 0.82 \times D_{\mathrm{H}\beta} + 0.80 \times R4570,
\label{du_recalibrated}
\end{equation}

where $D_{\mathrm{H}\beta} = \mathrm{FWHM}/\sigma_{\mathrm{H}\beta}$, and $\sigma_{\mathrm{H}\beta}$ is the dispersion of the broad \hb\ line. Using this relation, we estimate the Eddington ratio for each NLSy1 source based on its measured \feii strength and \hb profile shape. We note that this relation was derived from a sample of only 64 reverberation-mapped AGNs, and thus may be affected by low-number statistics. However, anchoring the calibration to reverberation mapping-based black hole masses reduces systematic biases relative to single-epoch estimates.

Finally, we also fit the \citet{2016ApJ...818L..14D} multivariate relation between $\log_{10}(R_{\mathrm{Edd}})$, $D_{\mathrm{H}\beta}$, and R4570 for our BLSy1 sample of 1143 sources. The $D_{\mathrm{H}\beta}$ values were measured using PyQSOFit \citep{2018ascl.soft09008G}. The resulting correlation, $\log_{10}(R_{\mathrm{Edd}}) = (-0.33 \pm 0.04) - (0.52 \pm 0.02)\, D_{\mathrm{H}\beta} + (0.59 \pm 0.05)\, R4570$ is consistent with the equation~\ref{du_recalibrated}. Fig.~\ref{fig:du_recalibrated} compares the catalog $R_{\mathrm{Edd}}$ values with those recalibrated using this relation for the BLSy1 sample. The dashed line indicates the one-to-one correspondence, showing good agreement with modest scatter. Nonetheless, our calibration depends on  $R_{\mathrm{Edd}}$ values derived from single-epoch virial black hole masses and is therefore more susceptible to systematic biases than the \citet{2016ApJ...818L..14D} relation.

The $R_{\text{Edd}}$ values are then incorporated into equation~\ref{zhou}, which uses $\tau_{\text{DRW}}$, $R_{\text{Edd}}$, and rest-frame wavelength to compute the black hole masses. Our second approach, using the equation~\ref{du_recalibrated}, yields a median DRW-based black hole mass of $\log(M_{\mathrm{BH}}^{\mathrm{DRW}}/M_{\odot}) = 6.25 \pm 0.65$ for NLSy1s and $7.07 \pm 0.67$ for BLSy1s. Finally, our third approach, using the adapted \citet{2016ApJ...818L..14D} relation calibrated for our BLSy1 sample, yields $\log(M_{\mathrm{BH}}^{\mathrm{DRW}}/M_{\odot}) = 6.62 \pm 0.46$ for NLSy1s and $7.20 \pm 0.47$ for BLSy1s. This suggests that, regardless of the method adopted, the DRW-based masses for NLSy1s remain systematically lower than those for BLSy1s. Fig.~\ref{fig:zhou_distribution} shows the distribution of black hole masses ($\log M_{\mathrm{BH}}$) for the final sample of NLSy1 (red) and BLSy1 (black) sources obtained using the second approach (equation \ref{du_recalibrated}). The single-epoch mass distributions are shown as shaded histograms, with red and black shading representing the NLSy1 and BLSy1 samples, respectively. The figure clearly illustrates that, after correcting for Eddington ratio effects, NLSy1 galaxies occupy a lower black hole mass regime compared to BLSy1s. We also note that DRW-based masses from equation~\ref{du_recalibrated} are systematically lower than the single-epoch estimates, while those from equation~\ref{burke1} are systematically higher, indicating the presence of a residual zero-point offset in DRW-based mass predictions.
\section{Discussion}\label{5}
In this study, we measured the black hole masses of NLSy1 galaxies by modeling their optical variability within the framework of the DRW formalism. The widely held view that NLSy1s host relatively low-mass black holes is largely based on virial mass estimates, which rely on the FWHM of the H$\beta$ line and are therefore sensitive to orientation effects \citep[e.g.,][]{2004ApJ...606L..41G}. To mitigate these orientation-related biases, we applied the DRW method to model their optical variability using high-cadence $g$-band light curves from the ZTF survey. Our final sample consists of 1,141 NLSy1 and 1,143 BLSy1 galaxies, matched in the $L$–$z$ plane. To ensure robust constraints on variability timescales, we made use of both archival and forced photometry light curves from ZTF. The archival light curves generally provide higher SNR measurements, but are limited by source detection thresholds and become incomplete for fainter epochs. While the forced photometry recovers fluxes at fixed source positions even when the source is weakly detected in individual epochs, it provides many more data points and smoother light curve coverage, though at the cost of larger formal photometric errors. The denser sampling it offers is particularly valuable for constraining the DRW damping timescale, $\tau_{\mathrm{DRW}}$. Moreover, forced photometry also adds extra observing seasons, effectively extending the temporal baseline and further improving the robustness of the inferred timescales. We recommend using forced photometry, particularly for sources with small  $\tau_{DRW}$. Crucially, our main result is unchanged regardless of whether we use forced or archival photometry. Forced-photometry light curves primarily tighten the inferred $M_{\mathrm{BH}}^{\mathrm{DRW}}$
distributions by suppressing the extended low-mass tail.

Using the DRW modeling of these light curves, we obtained the two key DRW variability parameters, $\tau_{\mathrm{DRW}}$ and $\sigma$, explored their correlations with physical properties such as $M_{\mathrm{BH}}$, $R_{\mathrm{Edd}}$, and spectral diagnostics. This analysis revealed that DRW-based black hole mass estimates, when calibrated using the equation~\ref{burke1} under the assumption of a constant $R_{\mathrm{Edd}}$, are systematically overestimated for NLSy1s. The calibration yielded median masses of $\log(M_{\mathrm{BH}}^{\mathrm{DRW}}/M_\odot) = 7.94 \pm 0.38$ for NLSy1s and $7.97 \pm 0.36$ for BLSy1s, significantly exceeding the corresponding SE-based values of $6.81 \pm 0.38$ and $7.49 \pm 0.41$, respectively. While the DRW-based approach for BLSy1 yields broadly consistent results, with only a slight upward bias attributable to their relatively narrow $R_{\mathrm{Edd}}$ range (0.001–0.95), it substantially overestimates black hole masses in NLSy1s. This discrepancy arises from the calibration itself in equation~\ref{burke1}, which assumes a constant Eddington ratio across all sources, a condition that does not hold for NLSy1s. In our sample, NLSy1s span a much broader and higher range of $R_{\mathrm{Edd}}$ values (0.03–6.54), introducing substantial variation in the observed damping timescales. Because $\tau_{\text{DRW}}$ depends not just on black hole mass, but also on $R_{\text{Edd}}$ as suggested by accretion disk theory, ignoring these factors can make a black hole appear more massive, simply because its variability timescale is longer.

To address this discrepancy, we incorporated a multivariate calibration from equation~\ref{zhou} that explicitly includes $\tau_{\text{DRW}}$,  $R_{\text{Edd}}$ and rest-frame wavelength. Since $R_{\text{Edd}}$ is defined as the ratio of bolometric to Eddington luminosity, it is inherently linked to the black hole mass. Crucially, using the catalog SE, virial-based black hole masses to compute $R_{\text{Edd}}$
and then employing that $R_{\text{Edd}}$ in equation~\ref{zhou} to estimate $M_{\mathrm{BH}}$ introduces a cyclic bias. 
To reduce such biases, we explored the use of R4570 (the \feii/\hb flux ratio) as a more robust, inclination-independent proxy for $R_{\text{Edd}}$. We first tested the direct correlation between \feii strength, quantified by R4570, and $R_{\mathrm{Edd}}$ for BLSy1s, in line with earlier suggestions that R4570 traces accretion rate. However, in our BLSy1 sample, the simple linear R4570–$R_{\mathrm{Edd}}$ relation exhibits substantial intrinsic scatter and yields $R_{\text{Edd}}$ distributions that disagree with the \cite{2024MNRAS.527.7055P} catalog values.  
We therefore conclude that a simple linear dependence on R4570 does not work for reliable mass estimation; the $R_{\mathrm{Edd}}$-R4570 connection is inherently multivariate and requires additional spectral information.

To further refine the estimation, we incorporated broad line region (BLR) kinematic information through the H$\beta$ line shape parameter $D_{\mathrm{H}\beta}$. Applying the relation from equation~\ref{du_recalibrated}, which links $R_{\mathrm{Edd}}$ to both R4570 and $D_{\mathrm{H}\beta}$, we obtained even lower median masses of $\log(M_{\mathrm{BH}}^{\mathrm{DRW}}/M_{\odot}) = 6.25 \pm 0.65$ for NLSy1s and $7.07 \pm 0.67$ for BLSy1s. We also derived an independent recalibration using our own BLSy1 sample, based on the same parameters. This yielded median masses of $6.62 \pm 0.46$ for NLSy1s and $7.20 \pm 0.47$ for BLSy1s, consistent within the uncertainties reported by \citet{2021Sci...373..789B}. Both approaches produced similar results, and the mass estimates for BLSy1s remain systematically higher than those for NLSy1s. This persistent offset reinforces the physical distinction between the two populations and highlights the need to correct $\tau_{\mathrm{DRW}}$ for accretion-rate effects in high-$R_{\mathrm{Edd}}$ AGNs.
Further, the zero-point offsets of the corrected DRW-based masses from all methods relative to the SE estimates, which likely reflect differences in the calibration relations, suggest that further anchoring of DRW-based scalings to reverberation-mapped samples will be needed to resolve the offset. To improve this, future work should tie the DRW scaling relations more directly to reverberation-mapped masses so that a consistent zero-point can be established.

In our sample, we also identify 161 NLSy1 sources with absolute magnitudes brighter than $M_B = -23$, classifying them as narrow-line quasars. These objects tend to have higher black hole masses. However, even after excluding these quasars to isolate a more conservative sample of NLSy1 galaxies, we obtain a median DRW-based black hole mass of $\log(M_{\mathrm{BH}}^{\mathrm{DRW}}/M_{\odot}) = 6.22 \pm 0.69$, indicating a further downward shift in the mass distribution, as expected for a purer NLSy1 population. Additionally, our sample includes 16 radio-loud NLSy1 galaxies, a small subset that can host relativistic jets and, in principle, alter the optical continuum, variability amplitudes/timescales, and emission line properties. However, this subset is too small to drive the population trends reported here, and our conclusions remain unchanged.

For the initial sample of BLSy1 galaxies, we obtained a median single-epoch virial mass of $\log(M_{\mathrm{BH}}/M_\odot) = 7.49$. This value, itself, is somewhat lower than typically reported in the literature, likely reflecting the fact that our $L$–$z$ controlled sample was constructed to match the lower luminosities of the NLSy1s. As a result, the BLSy1 masses in our analysis carry a mild downward bias, which should be considered when comparing to studies based on luminosity-unrestricted BLSy1 samples. Furthermore, our sample also shows an inverse correlation between the variability amplitude and both luminosity and Fe II strength, consistent with the findings of \citet{2017ApJ...842...96R}. Taken together, our findings support a consistent interpretation: the relatively low variability amplitudes and longer damping timescales observed in NLSy1s are more likely driven by their higher accretion rates rather than by systematically larger black hole masses. By incorporating Eddington ratio corrections and adopting inclination-independent proxies such as R4570, we achieve more physically motivated and robust mass estimates. This approach effectively mitigates orientation-induced biases, particularly important given that our sample likely includes sources across a wide range of viewing angles. More broadly, these results underscore the diagnostic power of time-domain variability in constraining accretion physics and black hole demographics in active galaxies.

\section{Conclusion}
In this work, we used optical variability modeling to recalibrate black hole mass estimates for a large sample of NLSy1 and BLSy1 galaxies. By applying the DRW formalism to archival $g$-band light curves from the ZTF survey, we derived characteristic variability timescales and amplitudes for each source. Our analysis demonstrates that DRW–mass scaling relations only depend on blackhole masses, systematically overestimate black hole masses in NLSy1s, primarily because they neglect the impact of Eddington ratio on variability timescales.

To address this limitation, we implemented a multivariate relation that incorporates both the Eddington ratio and the rest-frame wavelength, resulting in more physically consistent mass estimates. Additionally, we employed the \feii strength (R4570) as an inclination-independent proxy for accretion rate, and further refined the calibration using the H$\beta$ line shape parameter $D_{\mathrm{H}\beta}$ which traces BLR kinematics.

Our findings suggest the interpretation that, as a population, NLSy1 galaxies harbor lower-mass, rapidly accreting black holes compared to their BLSy1 counterparts. More broadly, this study highlights the importance of accounting for accretion-driven variability effects in time-domain analyses and demonstrates the utility of DRW modeling as a powerful tool for probing AGN black hole demographics in large photometric surveys.

\section{Acknowledgments}
We thank the anonymous referee for the feedback which
has significantly helped to improve the paper. This work makes use of data from the Zwicky Transient Facility (ZTF). The ZTF forced-photometry service was funded under a grant from the Heising–Simons Foundation. Observations were obtained with the Samuel Oschin 48-inch and the 60-inch telescopes at Palomar Observatory as part of the ZTF project (IRSA 2022). ZTF is supported by the National Science Foundation under grants AST-1440341 and AST-2034437, together with a collaboration including Caltech, IPAC, the Weizmann Institute for Science, the Oskar Klein Center at Stockholm University, the University of Maryland, Deutsches Elektronen-Synchrotron and Humboldt University, the TANGO Consortium of Taiwan, the University of Wisconsin–Milwaukee, Trinity College Dublin, Lawrence Livermore National Laboratory, IN2P3, the University of Warwick, Ruhr University Bochum, and Northwestern University, as well as former partners the University of Washington, Los Alamos National Laboratory, and Lawrence Berkeley National Laboratory. Operations are carried out jointly by COO, IPAC, and UW.  We acknowledge support from the Department of Science and Technology, India - Science and Engineering Research Board (DST-SERB) in the form of a core research grant (CRG/2022/007884).

\section*{Data Availability}
All ZTF light-curve data used in this work are publicly available through the ZTF Data Release 23. Forced-photometry light curves can be obtained upon request from the ZTF archive. Spectroscopic data are publicly accessible via the SDSS archive. The reduced parameters underlying this study are provided in machine-readable ASCII/CSV format. 




\bibliographystyle{mnras}
\bibliography{cit} 

@ARTICLE{2024MNRAS.527.7055P,
       author = {{Paliya}, Vaidehi S. and {Stalin}, C.~S. and {Dom{\'\i}nguez}, Alberto and {Saikia}, D.~J.},
        title = "{Narrow-line Seyfert 1 galaxies in Sloan Digital Sky Survey: a new optical spectroscopic catalogue}",
      journal = {\mnras},
         year = 2024,
        month = jan,
       volume = {527},
       number = {3},
        pages = {7055-7069},
          doi = {10.1093/mnras/stad3650},
archivePrefix = {arXiv},
       eprint = {2311.13818},
 primaryClass = {astro-ph.GA},
       adsurl = {https://ui.adsabs.harvard.edu/abs/2024MNRAS.527.7055P}
}

@ARTICLE{2021Sci...373..789B,
       author = {{Burke}, Colin J. and {Shen}, Yue and {Blaes}, Omer and {Gammie}, Charles F. and {Horne}, Keith and {Jiang}, Yan-Fei and {Liu}, Xin and {McHardy}, Ian M. and {Morgan}, Christopher W. and {Scaringi}, Simone and {Yang}, Qian},
        title = "{A characteristic optical variability time scale in astrophysical accretion disks}",
      journal = {Science},
         year = 2021,
        month = aug,
       volume = {373},
       number = {6556},
        pages = {789-792},
          doi = {10.1126/science.abg9933},
archivePrefix = {arXiv},
       eprint = {2108.05389},
 primaryClass = {astro-ph.GA},
       adsurl = {https://ui.adsabs.harvard.edu/abs/2021Sci...373..789B}
}

@ARTICLE{2017A&A...597A.128K,
       author = {{Koz{\l}owski}, Szymon},
        title = "{Limitations on the recovery of the true AGN variability parameters using damped random walk modeling}",
      journal = {\aap},
         year = 2017,
        month = jan,
       volume = {597},
          eid = {A128},
        pages = {A128},
          doi = {10.1051/0004-6361/201629890},
archivePrefix = {arXiv},
       eprint = {1611.08248},
 primaryClass = {astro-ph.GA},
       adsurl = {https://ui.adsabs.harvard.edu/abs/2017A&A...597A.128K}
}

@ARTICLE{2001A&A...372..730V,
       author = {{V{\'e}ron-Cetty}, M. -P. and {V{\'e}ron}, P. and {Gon{\c{c}}alves}, A.~C.},
        title = "{A spectrophotometric atlas of Narrow-Line Seyfert 1 galaxies}",
      journal = {\aap},
         year = 2001,
        month = jun,
       volume = {372},
        pages = {730-754},
          doi = {10.1051/0004-6361:20010489},
archivePrefix = {arXiv},
       eprint = {astro-ph/0104151},
 primaryClass = {astro-ph},
       adsurl = {https://ui.adsabs.harvard.edu/abs/2001A&A...372..730V}
}

@ARTICLE{2004ApJ...606L..41G,
       author = {{Grupe}, Dirk and {Mathur}, Smita},
        title = "{M$_{BH}$-{\ensuremath{\sigma}} Relation for a Complete Sample of Soft X-Ray-selected Active Galactic Nuclei}",
      journal = {\apjl},
         year = 2004,
        month = may,
       volume = {606},
       number = {1},
        pages = {L41-L44},
          doi = {10.1086/420975},
archivePrefix = {arXiv},
       eprint = {astro-ph/0312390},
 primaryClass = {astro-ph},
       adsurl = {https://ui.adsabs.harvard.edu/abs/2004ApJ...606L..41G}
}

@ARTICLE{2016ApJ...819..121P,
       author = {{Paliya}, Vaidehi S. and {Rajput}, Bhoomika and {Stalin}, C.~S. and {Pandey}, S.~B.},
        title = "{Broadband Observations of the Gamma-Ray Emitting Narrow Line Seyfert 1 Galaxy SBS 0846+513}",
      journal = {\apj},
         year = 2016,
        month = mar,
       volume = {819},
       number = {2},
          eid = {121},
        pages = {121},
          doi = {10.3847/0004-637X/819/2/121},
archivePrefix = {arXiv},
       eprint = {1603.01534},
 primaryClass = {astro-ph.HE},
       adsurl = {https://ui.adsabs.harvard.edu/abs/2016ApJ...819..121P}
}

@ARTICLE{2013ApJ...765..106Z,
       author = {{Zu}, Ying and {Kochanek}, C.~S. and {Koz{\l}owski}, Szymon and {Udalski}, Andrzej},
        title = "{Is Quasar Optical Variability a Damped Random Walk?}",
      journal = {\apj},
         year = 2013,
        month = mar,
       volume = {765},
       number = {2},
          eid = {106},
        pages = {106},
          doi = {10.1088/0004-637X/765/2/106},
archivePrefix = {arXiv},
       eprint = {1202.3783},
 primaryClass = {astro-ph.CO},
       adsurl = {https://ui.adsabs.harvard.edu/abs/2013ApJ...765..106Z}
}

@ARTICLE{2024ApJ...967L..18Z,
       author = {{Zhang}, Haoyang and {Yang}, Shenbang and {Dai}, Benzhong},
        title = "{Discovering the Mass-Scaled Damping Timescale from Microquasars to Blazars}",
      journal = {\apjl},
         year = 2024,
        month = may,
       volume = {967},
       number = {1},
          eid = {L18},
        pages = {L18},
          doi = {10.3847/2041-8213/ad488d},
archivePrefix = {arXiv},
       eprint = {2405.05575},
 primaryClass = {astro-ph.HE},
       adsurl = {https://ui.adsabs.harvard.edu/abs/2024ApJ...967L..18Z}
}

@ARTICLE{2018MNRAS.480...96W,
       author = {{Williams}, James K. and {Gliozzi}, Mario and {Rudzinsky}, Ross V.},
        title = "{Are narrow-line Seyfert 1 galaxies highly accreting low-M$_{BH}$ AGNs?}",
      journal = {\mnras},
         year = 2018,
        month = oct,
       volume = {480},
       number = {1},
        pages = {96-107},
          doi = {10.1093/mnras/sty1868},
       adsurl = {https://ui.adsabs.harvard.edu/abs/2018MNRAS.480...96W}
}

@ARTICLE{2009ApJ...699..976A,
       author = {{Abdo}, A.~A. and {Ackermann}, M. and {Ajello}, M. and {Axelsson}, M. and {Baldini}, L. and {Ballet}, J. and {Barbiellini}, G. and {Bastieri}, D. and {Battelino}, M. and {Baughman}, B.~M. and {Bechtol}, K. and {Bellazzini}, R. and {Bloom}, E.~D. and {Bonamente}, E. and {Borgland}, A.~W. and {Bregeon}, J. and {Brez}, A. and {Brigida}, M. and {Bruel}, P. and {Caliandro}, G.~A. and {Cameron}, R.~A. and {Caraveo}, P.~A. and {Casandjian}, J.~M. and {Cavazzuti}, E. and {Cecchi}, C. and {Chekhtman}, A. and {Cheung}, C.~C. and {Chiang}, J. and {Ciprini}, S. and {Claus}, R. and {Cohen-Tanugi}, J. and {Collmar}, W. and {Conrad}, J. and {Costamante}, L. and {Dermer}, C.~D. and {de Angelis}, A. and {de Palma}, F. and {Digel}, S.~W. and {Silva}, E. do Couto e. and {Drell}, P.~S. and {Dubois}, R. and {Dumora}, D. and {Farnier}, C. and {Favuzzi}, C. and {Focke}, W.~B. and {Foschini}, L. and {Frailis}, M. and {Fuhrmann}, L. and {Fukazawa}, Y. and {Funk}, S. and {Fusco}, P. and {Gargano}, F. and {Gehrels}, N. and {Germani}, S. and {Giebels}, B. and {Giglietto}, N. and {Giordano}, F. and {Giroletti}, M. and {Glanzman}, T. and {Grenier}, I.~A. and {Grondin}, M. -H. and {Grove}, J.~E. and {Guillemot}, L. and {Guiriec}, S. and {Hanabata}, Y. and {Harding}, A.~K. and {Hartman}, R.~C. and {Hayashida}, M. and {Hays}, E. and {Hughes}, R.~E. and {J{\'o}hannesson}, G. and {Johnson}, A.~S. and {Johnson}, R.~P. and {Johnson}, W.~N. and {Kamae}, T. and {Katagiri}, H. and {Kataoka}, J. and {Kerr}, M. and {Kn{\"o}dlseder}, J. and {Kuehn}, F. and {Kuss}, M. and {Lande}, J. and {Latronico}, L. and {Lemoine-Goumard}, M. and {Longo}, F. and {Loparco}, F. and {Lott}, B. and {Lovellette}, M.~N. and {Lubrano}, P. and {Madejski}, G.~M. and {Makeev}, A. and {Max-Moerbeck}, W. and {Mazziotta}, M.~N. and {McConville}, W. and {McEnery}, J.~E. and {Meurer}, C. and {Michelson}, P.~F. and {Mitthumsiri}, W. and {Mizuno}, T. and {Monte}, C. and {Monzani}, M.~E. and {Morselli}, A. and {Moskalenko}, I.~V. and {Murgia}, S. and {Nolan}, P.~L. and {Norris}, J.~P. and {Nuss}, E. and {Ohsugi}, T. and {Omodei}, N. and {Orlando}, E. and {Ormes}, J.~F. and {Paneque}, D. and {Panetta}, J.~H. and {Parent}, D. and {Pavlidou}, V. and {Pearson}, T.~J. and {Pepe}, M. and {Pesce-Rollins}, M. and {Piron}, F. and {Porter}, T.~A. and {Rain{\`o}}, S. and {Rando}, R. and {Razzano}, M. and {Readhead}, A. and {Reimer}, A. and {Reimer}, O. and {Reposeur}, T. and {Richards}, J.~L. and {Ritz}, S. and {Rodriguez}, A.~Y. and {Romani}, R.~W. and {Ryde}, F. and {Sadrozinski}, H.~F. -W. and {Sambruna}, R. and {Sanchez}, D. and {Sander}, A. and {Parkinson}, P.~M. Saz and {Scargle}, J.~D. and {Schalk}, T.~L. and {Sgr{\`o}}, C. and {Smith}, D.~A. and {Spandre}, G. and {Spinelli}, P. and {Starck}, J. -L. and {Stevenson}, M. and {Strickman}, M.~S. and {Suson}, D.~J. and {Tagliaferri}, G. and {Takahashi}, H. and {Tanaka}, T. and {Thayer}, J.~G. and {Thompson}, D.~J. and {Tibaldo}, L. and {Tibolla}, O. and {Torres}, D.~F. and {Tosti}, G. and {Tramacere}, A. and {Uchiyama}, Y. and {Usher}, T.~L. and {Vilchez}, N. and {Vitale}, V. and {Waite}, A.~P. and {Winer}, B.~L. and {Wood}, K.~S. and {Ylinen}, T. and {Zensus}, J.~A. and {Ziegler}, M. and {Fermi/LAT Collaboration} and {Ghisellini}, G. and {Maraschi}, L. and {Tavecchio}, F. and {Angelakis}, E.},
        title = "{Fermi/Large Area Telescope Discovery of Gamma-Ray Emission from a Relativistic Jet in the Narrow-Line Quasar PMN J0948+0022}",
      journal = {\apj},
         year = 2009,
        month = jul,
       volume = {699},
       number = {2},
        pages = {976-984},
          doi = {10.1088/0004-637X/699/2/976},
archivePrefix = {arXiv},
       eprint = {0905.4558},
 primaryClass = {astro-ph.HE},
       adsurl = {https://ui.adsabs.harvard.edu/abs/2009ApJ...699..976A}
}

@PROCEEDINGS{2011nlsg.confE....F,
        title = "{Narrow-Line Seyfert 1 Galaxies and their place in the Universe}",
    booktitle = {Narrow-Line Seyfert 1 Galaxies and their Place in the Universe},
         year = 2011,
       editor = {{Foschini}, L. and {Colpi}, M. and {Gallo}, L. and {Grupe}, D. and {Komossa}, S. and {Leighly}, K. and {Mathur}, S.},
        month = jan,
       adsurl = {https://ui.adsabs.harvard.edu/abs/2011nlsg.confE....F}
}

@ARTICLE{2012MNRAS.426..317D,
       author = {{D'Ammando}, F. and {Orienti}, M. and {Finke}, J. and {Raiteri}, C.~M. and {Angelakis}, E. and {Fuhrmann}, L. and {Giroletti}, M. and {Hovatta}, T. and {Max-Moerbeck}, W. and {Perkins}, J.~S. and {Readhead}, A.~C.~S. and {Richards}, J.~L. and {Stawarz}, {\L}. and {Donato}, D.},
        title = "{SBS 0846+513: a new {\ensuremath{\gamma}}-ray-emitting narrow-line Seyfert 1 galaxy}",
      journal = {\mnras},
         year = 2012,
        month = oct,
       volume = {426},
       number = {1},
        pages = {317-329},
          doi = {10.1111/j.1365-2966.2012.21707.x},
archivePrefix = {arXiv},
       eprint = {1207.3092},
 primaryClass = {astro-ph.HE},
       adsurl = {https://ui.adsabs.harvard.edu/abs/2012MNRAS.426..317D}
}

@ARTICLE{2018ApJ...853L...2P,
       author = {{Paliya}, Vaidehi S. and {Ajello}, M. and {Rakshit}, S. and {Mandal}, Amit Kumar and {Stalin}, C.~S. and {Kaur}, A. and {Hartmann}, D.},
        title = "{Gamma-Ray-emitting Narrow-line Seyfert 1 Galaxies in the Sloan Digital Sky Survey}",
      journal = {\apjl},
         year = 2018,
        month = jan,
       volume = {853},
       number = {1},
          eid = {L2},
        pages = {L2},
          doi = {10.3847/2041-8213/aaa5ab},
archivePrefix = {arXiv},
       eprint = {1801.01905},
 primaryClass = {astro-ph.HE},
       adsurl = {https://ui.adsabs.harvard.edu/abs/2018ApJ...853L...2P}
}

@ARTICLE{1996A&A...309...81W,
       author = {{Wang}, T. and {Brinkmann}, W. and {Bergeron}, J.},
        title = "{X-ray properties of active galactic nuclei with optical FeII emission.}",
      journal = {\aap},
         year = 1996,
        month = may,
       volume = {309},
        pages = {81-96},
       adsurl = {https://ui.adsabs.harvard.edu/abs/1996A&A...309...81W}
}

@ARTICLE{2019PASP..131a8003M,
       author = {{Masci}, Frank J. and {Laher}, Russ R. and {Rusholme}, Ben and {Shupe}, David L. and {Groom}, Steven and {Surace}, Jason and {Jackson}, Edward and {Monkewitz}, Serge and {Beck}, Ron and {Flynn}, David and {Terek}, Scott and {Landry}, Walter and {Hacopians}, Eugean and {Desai}, Vandana and {Howell}, Justin and {Brooke}, Tim and {Imel}, David and {Wachter}, Stefanie and {Ye}, Quan-Zhi and {Lin}, Hsing-Wen and {Cenko}, S. Bradley and {Cunningham}, Virginia and {Rebbapragada}, Umaa and {Bue}, Brian and {Miller}, Adam A. and {Mahabal}, Ashish and {Bellm}, Eric C. and {Patterson}, Maria T. and {Juri{\'c}}, Mario and {Golkhou}, V. Zach and {Ofek}, Eran O. and {Walters}, Richard and {Graham}, Matthew and {Kasliwal}, Mansi M. and {Dekany}, Richard G. and {Kupfer}, Thomas and {Burdge}, Kevin and {Cannella}, Christopher B. and {Barlow}, Tom and {Van Sistine}, Angela and {Giomi}, Matteo and {Fremling}, Christoffer and {Blagorodnova}, Nadejda and {Levitan}, David and {Riddle}, Reed and {Smith}, Roger M. and {Helou}, George and {Prince}, Thomas A. and {Kulkarni}, Shrinivas R.},
        title = "{The Zwicky Transient Facility: Data Processing, Products, and Archive}",
      journal = {\pasp},
         year = 2019,
        month = jan,
       volume = {131},
       number = {995},
        pages = {018003},
          doi = {10.1088/1538-3873/aae8ac},
archivePrefix = {arXiv},
       eprint = {1902.01872},
 primaryClass = {astro-ph.IM},
       adsurl = {https://ui.adsabs.harvard.edu/abs/2019PASP..131a8003M}
}

@ARTICLE{2017AJ....154..220F,
       author = {{Foreman-Mackey}, Daniel and {Agol}, Eric and {Ambikasaran}, Sivaram and {Angus}, Ruth},
        title = "{Fast and Scalable Gaussian Process Modeling with Applications to Astronomical Time Series}",
      journal = {\aj},
         year = 2017,
        month = dec,
       volume = {154},
       number = {6},
          eid = {220},
        pages = {220},
          doi = {10.3847/1538-3881/aa9332},
archivePrefix = {arXiv},
       eprint = {1703.09710},
 primaryClass = {astro-ph.IM},
       adsurl = {https://ui.adsabs.harvard.edu/abs/2017AJ....154..220F}
}

@ARTICLE{2013PASP..125..306F,
       author = {{Foreman-Mackey}, Daniel and {Hogg}, David W. and {Lang}, Dustin and {Goodman}, Jonathan},
        title = "{emcee: The MCMC Hammer}",
      journal = {\pasp},
         year = 2013,
        month = mar,
       volume = {125},
       number = {925},
        pages = {306},
          doi = {10.1086/670067},
archivePrefix = {arXiv},
       eprint = {1202.3665},
 primaryClass = {astro-ph.IM},
       adsurl = {https://ui.adsabs.harvard.edu/abs/2013PASP..125..306F}
}

@ARTICLE{2024ApJ...966....8Z,
       author = {{Zhou}, Shuying and {Sun}, Mouyuan and {Cai}, Zhen-Yi and {Ren}, Guowei and {Wang}, Jun-Xian and {Xue}, Yongquan},
        title = "{How Long Will the Quasar UV/Optical Flickering Be Damped?}",
      journal = {\apj},
         year = 2024,
        month = may,
       volume = {966},
       number = {1},
          eid = {8},
        pages = {8},
          doi = {10.3847/1538-4357/ad2fbc},
archivePrefix = {arXiv},
       eprint = {2403.01691},
 primaryClass = {astro-ph.HE},
       adsurl = {https://ui.adsabs.harvard.edu/abs/2024ApJ...966....8Z}
}

@ARTICLE{1985ApJ...297..166O,
       author = {{Osterbrock}, D.~E. and {Pogge}, R.~W.},
        title = "{The spectra of narrow-line Seyfert 1 galaxies.}",
      journal = {\apj},
         year = 1985,
        month = oct,
       volume = {297},
        pages = {166-176},
          doi = {10.1086/163513},
       adsurl = {https://ui.adsabs.harvard.edu/abs/1985ApJ...297..166O}
}

@ARTICLE{1989ApJ...342..224G,
       author = {{Goodrich}, Robert W.},
        title = "{Spectropolarimetry of ``Narrow-Line'' Seyfert 1 Galaxies}",
      journal = {\apj},
         year = 1989,
        month = jul,
       volume = {342},
        pages = {224},
          doi = {10.1086/167586},
       adsurl = {https://ui.adsabs.harvard.edu/abs/1989ApJ...342..224G}
}

@ARTICLE{1996A&A...305...53B,
       author = {{Boller}, T. and {Brandt}, W.~N. and {Fink}, H.},
        title = "{Soft X-ray properties of narrow-line Seyfert 1 galaxies.}",
      journal = {\aap},
         year = 1996,
        month = jan,
       volume = {305},
        pages = {53},
          doi = {10.48550/arXiv.astro-ph/9504093},
archivePrefix = {arXiv},
       eprint = {astro-ph/9504093},
 primaryClass = {astro-ph},
       adsurl = {https://ui.adsabs.harvard.edu/abs/1996A&A...305...53B}
}

@ARTICLE{1999ApJS..125..317L,
       author = {{Leighly}, Karen M.},
        title = "{A Comprehensive Spectral and Variability Study of Narrow-Line Seyfert 1 Galaxies Observed by ASCA. II. Spectral Analysis and Correlations}",
      journal = {\apjs},
         year = 1999,
        month = dec,
       volume = {125},
       number = {2},
        pages = {317-348},
          doi = {10.1086/313287},
archivePrefix = {arXiv},
       eprint = {astro-ph/9907295},
 primaryClass = {astro-ph},
       adsurl = {https://ui.adsabs.harvard.edu/abs/1999ApJS..125..317L}
}

@ARTICLE{2021iSci...24j2557C,
       author = {{Cackett}, Edward M. and {Bentz}, Misty C. and {Kara}, Erin},
        title = "{Reverberation mapping of active galactic nuclei: from X-ray corona to dusty torus}",
      journal = {iScience},
         year = 2021,
        month = jun,
       volume = {24},
       number = {6},
        pages = {102557},
          doi = {10.1016/j.isci.2021.102557},
archivePrefix = {arXiv},
       eprint = {2105.06926},
 primaryClass = {astro-ph.GA},
       adsurl = {https://ui.adsabs.harvard.edu/abs/2021iSci...24j2557C}
}

@ARTICLE{2016MNRAS.458L..69B,
       author = {{Baldi}, Ranieri D. and {Capetti}, Alessandro and {Robinson}, Andrew and {Laor}, Ari and {Behar}, Ehud},
        title = "{Radio-loud Narrow Line Seyfert 1 under a different perspective: a revised black hole mass estimate from optical spectropolarimetry}",
      journal = {\mnras},
         year = 2016,
        month = may,
       volume = {458},
       number = {1},
        pages = {L69-L73},
          doi = {10.1093/mnrasl/slw019},
archivePrefix = {arXiv},
       eprint = {1602.02783},
 primaryClass = {astro-ph.GA},
       adsurl = {https://ui.adsabs.harvard.edu/abs/2016MNRAS.458L..69B}
}

@ARTICLE{2013MNRAS.431..210C,
       author = {{Calderone}, G. and {Ghisellini}, G. and {Colpi}, M. and {Dotti}, M.},
        title = "{Black hole mass estimate for a sample of radio-loud narrow-line Seyfert 1 galaxies}",
      journal = {\mnras},
         year = 2013,
        month = may,
       volume = {431},
       number = {1},
        pages = {210-239},
          doi = {10.1093/mnras/stt157},
archivePrefix = {arXiv},
       eprint = {1212.1181},
 primaryClass = {astro-ph.CO},
       adsurl = {https://ui.adsabs.harvard.edu/abs/2013MNRAS.431..210C}
}

@ARTICLE{2019ApJ...881L..24V,
       author = {{Viswanath}, Gayathri and {Stalin}, C.~S. and {Rakshit}, Suvendu and {Kurian}, Kshama S. and {Ujjwal}, K. and {Gudennavar}, Shivappa B. and {Kartha}, Sreeja S.},
        title = "{Are Narrow-line Seyfert 1 Galaxies Powered by Low-mass Black Holes?}",
      journal = {\apjl},
         year = 2019,
        month = aug,
       volume = {881},
       number = {1},
          eid = {L24},
        pages = {L24},
          doi = {10.3847/2041-8213/ab365e},
archivePrefix = {arXiv},
       eprint = {1907.02683},
 primaryClass = {astro-ph.GA},
       adsurl = {https://ui.adsabs.harvard.edu/abs/2019ApJ...881L..24V}
}

@ARTICLE{2015MNRAS.446.3427C,
       author = {{Capellupo}, D.~M. and {Netzer}, H. and {Lira}, P. and {Trakhtenbrot}, B. and {Mej{\'\i}a-Restrepo}, Juli{\'a}n},
        title = "{Active galactic nuclei at z {\ensuremath{\sim}} 1.5 - I. Spectral energy distribution and accretion discs}",
      journal = {\mnras},
         year = 2015,
        month = feb,
       volume = {446},
       number = {4},
        pages = {3427-3446},
          doi = {10.1093/mnras/stu2266},
archivePrefix = {arXiv},
       eprint = {1410.8137},
 primaryClass = {astro-ph.GA},
       adsurl = {https://ui.adsabs.harvard.edu/abs/2015MNRAS.446.3427C}
}

@ARTICLE{2016MNRAS.460..212C,
       author = {{Capellupo}, D.~M. and {Netzer}, H. and {Lira}, P. and {Trakhtenbrot}, B. and {Mej{\'\i}a-Restrepo}, J.},
        title = "{Active galactic nuclei at z {\ensuremath{\sim}} 1.5 - III. Accretion discs and black hole spin}",
      journal = {\mnras},
         year = 2016,
        month = jul,
       volume = {460},
       number = {1},
        pages = {212-226},
          doi = {10.1093/mnras/stw937},
archivePrefix = {arXiv},
       eprint = {1604.05310},
 primaryClass = {astro-ph.GA},
       adsurl = {https://ui.adsabs.harvard.edu/abs/2016MNRAS.460..212C}
}

@ARTICLE{2018NatAs...2...63M,
       author = {{Mej{\'\i}a-Restrepo}, J.~E. and {Lira}, P. and {Netzer}, H. and {Trakhtenbrot}, B. and {Capellupo}, D.~M.},
        title = "{The effect of nuclear gas distribution on the mass determination of supermassive black holes}",
      journal = {Nature Astronomy},
         year = 2018,
        month = nov,
       volume = {2},
        pages = {63-68},
          doi = {10.1038/s41550-017-0305-z},
archivePrefix = {arXiv},
       eprint = {1709.05345},
 primaryClass = {astro-ph.GA},
       adsurl = {https://ui.adsabs.harvard.edu/abs/2018NatAs...2...63M}
}

@ARTICLE{2009ApJ...698..895K,
       author = {{Kelly}, Brandon C. and {Bechtold}, Jill and {Siemiginowska}, Aneta},
        title = "{Are the Variations in Quasar Optical Flux Driven by Thermal Fluctuations?}",
      journal = {\apj},
         year = 2009,
        month = jun,
       volume = {698},
       number = {1},
        pages = {895-910},
          doi = {10.1088/0004-637X/698/1/895},
archivePrefix = {arXiv},
       eprint = {0903.5315},
 primaryClass = {astro-ph.CO},
       adsurl = {https://ui.adsabs.harvard.edu/abs/2009ApJ...698..895K}
}

@ARTICLE{2010ApJ...708..927K,
       author = {{Koz{\l}owski}, Szymon and {Kochanek}, Christopher S. and {Udalski}, A. and {Wyrzykowski}, {\L}. and {Soszy{\'n}ski}, I. and {Szyma{\'n}ski}, M.~K. and {Kubiak}, M. and {Pietrzy{\'n}ski}, G. and {Szewczyk}, O. and {Ulaczyk}, K. and {Poleski}, R. and {OGLE Collaboration}},
        title = "{Quantifying Quasar Variability as Part of a General Approach to Classifying Continuously Varying Sources}",
      journal = {\apj},
         year = 2010,
        month = jan,
       volume = {708},
       number = {2},
        pages = {927-945},
          doi = {10.1088/0004-637X/708/2/927},
archivePrefix = {arXiv},
       eprint = {0909.1326},
 primaryClass = {astro-ph.CO},
       adsurl = {https://ui.adsabs.harvard.edu/abs/2010ApJ...708..927K}
}

@ARTICLE{2010ApJ...721.1014M,
       author = {{MacLeod}, C.~L. and {Ivezi{\'c}}, {\v{Z}}. and {Kochanek}, C.~S. and {Koz{\l}owski}, S. and {Kelly}, B. and {Bullock}, E. and {Kimball}, A. and {Sesar}, B. and {Westman}, D. and {Brooks}, K. and {Gibson}, R. and {Becker}, A.~C. and {de Vries}, W.~H.},
        title = "{Modeling the Time Variability of SDSS Stripe 82 Quasars as a Damped Random Walk}",
      journal = {\apj},
         year = 2010,
        month = oct,
       volume = {721},
       number = {2},
        pages = {1014-1033},
          doi = {10.1088/0004-637X/721/2/1014},
archivePrefix = {arXiv},
       eprint = {1004.0276},
 primaryClass = {astro-ph.CO},
       adsurl = {https://ui.adsabs.harvard.edu/abs/2010ApJ...721.1014M}
}

@ARTICLE{2001ApJ...555..775C,
       author = {{Collier}, Stefan and {Peterson}, Bradley M.},
        title = "{Characteristic Ultraviolet/Optical Timescales in Active Galactic Nuclei}",
      journal = {\apj},
         year = 2001,
        month = jul,
       volume = {555},
       number = {2},
        pages = {775-785},
          doi = {10.1086/321517},
       adsurl = {https://ui.adsabs.harvard.edu/abs/2001ApJ...555..775C}
}

@ARTICLE{2016A&A...585A.129S,
       author = {{Simm}, T. and {Salvato}, M. and {Saglia}, R. and {Ponti}, G. and {Lanzuisi}, G. and {Trakhtenbrot}, B. and {Nandra}, K. and {Bender}, R.},
        title = "{Pan-STARRS1 variability of XMM-COSMOS AGN. II. Physical correlations and power spectrum analysis}",
      journal = {\aap},
         year = 2016,
        month = jan,
       volume = {585},
          eid = {A129},
        pages = {A129},
          doi = {10.1051/0004-6361/201527353},
archivePrefix = {arXiv},
       eprint = {1510.06737},
 primaryClass = {astro-ph.GA},
       adsurl = {https://ui.adsabs.harvard.edu/abs/2016A&A...585A.129S}
}

@ARTICLE{2015SciA....1E0686S,
       author = {{Scaringi}, S. and {Maccarone}, T.~J. and {Kording}, E. and {Knigge}, C. and {Vaughan}, S. and {Marsh}, T.~R. and {Aranzana}, E. and {Dhillon}, V.~S. and {Barros}, S.~C.~C.},
        title = "{Accretion-induced variability links young stellar objects, white dwarfs, and black holes}",
      journal = {Science Advances},
         year = 2015,
        month = oct,
       volume = {1},
       number = {9},
        pages = {e1500686-e1500686},
          doi = {10.1126/sciadv.1500686},
archivePrefix = {arXiv},
       eprint = {1510.02471},
 primaryClass = {astro-ph.HE},
       adsurl = {https://ui.adsabs.harvard.edu/abs/2015SciA....1E0686S}
}

@ARTICLE{1973A&A....24..337S,
       author = {{Shakura}, N.~I. and {Sunyaev}, R.~A.},
        title = "{Black holes in binary systems. Observational appearance.}",
      journal = {\aap},
         year = 1973,
        month = jan,
       volume = {24},
        pages = {337-355},
       adsurl = {https://ui.adsabs.harvard.edu/abs/1973A&A....24..337S}
}

@ARTICLE{2016ApJ...818L..14D,
       author = {{Du}, Pu and {Wang}, Jian-Min and {Hu}, Chen and {Ho}, Luis C. and {Li}, Yan-Rong and {Bai}, Jin-Ming},
        title = "{The Fundamental Plane of the Broad-line Region in Active Galactic Nuclei}",
      journal = {\apjl},
         year = 2016,
        month = feb,
       volume = {818},
       number = {1},
          eid = {L14},
        pages = {L14},
          doi = {10.3847/2041-8205/818/1/L14},
archivePrefix = {arXiv},
       eprint = {1601.01391},
 primaryClass = {astro-ph.GA},
       adsurl = {https://ui.adsabs.harvard.edu/abs/2016ApJ...818L..14D}
}

@ARTICLE{2019PASP..131a8002B,
       author = {{Bellm}, Eric C. and {Kulkarni}, Shrinivas R. and {Graham}, Matthew J. and {Dekany}, Richard and {Smith}, Roger M. and {Riddle}, Reed and {Masci}, Frank J. and {Helou}, George and {Prince}, Thomas A. and {Adams}, Scott M. and {Barbarino}, C. and {Barlow}, Tom and {Bauer}, James and {Beck}, Ron and {Belicki}, Justin and {Biswas}, Rahul and {Blagorodnova}, Nadejda and {Bodewits}, Dennis and {Bolin}, Bryce and {Brinnel}, Valery and {Brooke}, Tim and {Bue}, Brian and {Bulla}, Mattia and {Burruss}, Rick and {Cenko}, S. Bradley and {Chang}, Chan-Kao and {Connolly}, Andrew and {Coughlin}, Michael and {Cromer}, John and {Cunningham}, Virginia and {De}, Kishalay and {Delacroix}, Alex and {Desai}, Vandana and {Duev}, Dmitry A. and {Eadie}, Gwendolyn and {Farnham}, Tony L. and {Feeney}, Michael and {Feindt}, Ulrich and {Flynn}, David and {Franckowiak}, Anna and {Frederick}, S. and {Fremling}, C. and {Gal-Yam}, Avishay and {Gezari}, Suvi and {Giomi}, Matteo and {Goldstein}, Daniel A. and {Golkhou}, V. Zach and {Goobar}, Ariel and {Groom}, Steven and {Hacopians}, Eugean and {Hale}, David and {Henning}, John and {Ho}, Anna Y.~Q. and {Hover}, David and {Howell}, Justin and {Hung}, Tiara and {Huppenkothen}, Daniela and {Imel}, David and {Ip}, Wing-Huen and {Ivezi{\'c}}, {\v{Z}}eljko and {Jackson}, Edward and {Jones}, Lynne and {Juric}, Mario and {Kasliwal}, Mansi M. and {Kaspi}, S. and {Kaye}, Stephen and {Kelley}, Michael S.~P. and {Kowalski}, Marek and {Kramer}, Emily and {Kupfer}, Thomas and {Landry}, Walter and {Laher}, Russ R. and {Lee}, Chien-De and {Lin}, Hsing Wen and {Lin}, Zhong-Yi and {Lunnan}, Ragnhild and {Giomi}, Matteo and {Mahabal}, Ashish and {Mao}, Peter and {Miller}, Adam A. and {Monkewitz}, Serge and {Murphy}, Patrick and {Ngeow}, Chow-Choong and {Nordin}, Jakob and {Nugent}, Peter and {Ofek}, Eran and {Patterson}, Maria T. and {Penprase}, Bryan and {Porter}, Michael and {Rauch}, Ludwig and {Rebbapragada}, Umaa and {Reiley}, Dan and {Rigault}, Mickael and {Rodriguez}, Hector and {van Roestel}, Jan and {Rusholme}, Ben and {van Santen}, Jakob and {Schulze}, S. and {Shupe}, David L. and {Singer}, Leo P. and {Soumagnac}, Maayane T. and {Stein}, Robert and {Surace}, Jason and {Sollerman}, Jesper and {Szkody}, Paula and {Taddia}, F. and {Terek}, Scott and {Van Sistine}, Angela and {van Velzen}, Sjoert and {Vestrand}, W. Thomas and {Walters}, Richard and {Ward}, Charlotte and {Ye}, Quan-Zhi and {Yu}, Po-Chieh and {Yan}, Lin and {Zolkower}, Jeffry},
        title = "{The Zwicky Transient Facility: System Overview, Performance, and First Results}",
      journal = {\pasp},
         year = 2019,
        month = jan,
       volume = {131},
       number = {995},
        pages = {018002},
          doi = {10.1088/1538-3873/aaecbe},
archivePrefix = {arXiv},
       eprint = {1902.01932},
 primaryClass = {astro-ph.IM},
       adsurl = {https://ui.adsabs.harvard.edu/abs/2019PASP..131a8002B}
}

@ARTICLE{2019PASP..131g8001G,
       author = {{Graham}, Matthew J. and {Kulkarni}, S.~R. and {Bellm}, Eric C. and {Adams}, Scott M. and {Barbarino}, Cristina and {Blagorodnova}, Nadejda and {Bodewits}, Dennis and {Bolin}, Bryce and {Brady}, Patrick R. and {Cenko}, S. Bradley and {Chang}, Chan-Kao and {Coughlin}, Michael W. and {De}, Kishalay and {Eadie}, Gwendolyn and {Farnham}, Tony L. and {Feindt}, Ulrich and {Franckowiak}, Anna and {Fremling}, Christoffer and {Gezari}, Suvi and {Ghosh}, Shaon and {Goldstein}, Daniel A. and {Golkhou}, V. Zach and {Goobar}, Ariel and {Ho}, Anna Y.~Q. and {Huppenkothen}, Daniela and {Ivezi{\'c}}, {\v{Z}}eljko and {Jones}, R. Lynne and {Juric}, Mario and {Kaplan}, David L. and {Kasliwal}, Mansi M. and {Kelley}, Michael S.~P. and {Kupfer}, Thomas and {Lee}, Chien-De and {Lin}, Hsing Wen and {Lunnan}, Ragnhild and {Mahabal}, Ashish A. and {Miller}, Adam A. and {Ngeow}, Chow-Choong and {Nugent}, Peter and {Ofek}, Eran O. and {Prince}, Thomas A. and {Rauch}, Ludwig and {van Roestel}, Jan and {Schulze}, Steve and {Singer}, Leo P. and {Sollerman}, Jesper and {Taddia}, Francesco and {Yan}, Lin and {Ye}, Quan-Zhi and {Yu}, Po-Chieh and {Barlow}, Tom and {Bauer}, James and {Beck}, Ron and {Belicki}, Justin and {Biswas}, Rahul and {Brinnel}, Valery and {Brooke}, Tim and {Bue}, Brian and {Bulla}, Mattia and {Burruss}, Rick and {Connolly}, Andrew and {Cromer}, John and {Cunningham}, Virginia and {Dekany}, Richard and {Delacroix}, Alex and {Desai}, Vandana and {Duev}, Dmitry A. and {Feeney}, Michael and {Flynn}, David and {Frederick}, Sara and {Gal-Yam}, Avishay and {Giomi}, Matteo and {Groom}, Steven and {Hacopians}, Eugean and {Hale}, David and {Helou}, George and {Henning}, John and {Hover}, David and {Hillenbrand}, Lynne A. and {Howell}, Justin and {Hung}, Tiara and {Imel}, David and {Ip}, Wing-Huen and {Jackson}, Edward and {Kaspi}, Shai and {Kaye}, Stephen and {Kowalski}, Marek and {Kramer}, Emily and {Kuhn}, Michael and {Landry}, Walter and {Laher}, Russ R. and {Mao}, Peter and {Masci}, Frank J. and {Monkewitz}, Serge and {Murphy}, Patrick and {Nordin}, Jakob and {Patterson}, Maria T. and {Penprase}, Bryan and {Porter}, Michael and {Rebbapragada}, Umaa and {Reiley}, Dan and {Riddle}, Reed and {Rigault}, Mickael and {Rodriguez}, Hector and {Rusholme}, Ben and {van Santen}, Jakob and {Shupe}, David L. and {Smith}, Roger M. and {Soumagnac}, Maayane T. and {Stein}, Robert and {Surace}, Jason and {Szkody}, Paula and {Terek}, Scott and {Van Sistine}, Angela and {van Velzen}, Sjoert and {Vestrand}, W. Thomas and {Walters}, Richard and {Ward}, Charlotte and {Zhang}, Chaoran and {Zolkower}, Jeffry},
        title = "{The Zwicky Transient Facility: Science Objectives}",
      journal = {\pasp},
         year = 2019,
        month = jul,
       volume = {131},
       number = {1001},
        pages = {078001},
          doi = {10.1088/1538-3873/ab006c},
archivePrefix = {arXiv},
       eprint = {1902.01945},
 primaryClass = {astro-ph.IM},
       adsurl = {https://ui.adsabs.harvard.edu/abs/2019PASP..131g8001G}
}

@ARTICLE{2023arXiv230516279M,
       author = {{Masci}, Frank J. and {Laher}, Russ R. and {Rusholme}, Benjamin and {Shupe}, David and {Paladini}, Roberta and {Groom}, Steve and {Wold}, Avery and {Miller}, Adam A. and {Drake}, Andrew},
        title = "{A New Forced Photometry Service for the Zwicky Transient Facility}",
      journal = {arXiv e-prints},
         year = 2023,
        month = may,
          eid = {arXiv:2305.16279},
        pages = {arXiv:2305.16279},
          doi = {10.48550/arXiv.2305.16279},
archivePrefix = {arXiv},
       eprint = {2305.16279},
 primaryClass = {astro-ph.IM},
       adsurl = {https://ui.adsabs.harvard.edu/abs/2023arXiv230516279M}
}

@ARTICLE{2017ApJ...842...96R,
       author = {{Rakshit}, Suvendu and {Stalin}, C.~S.},
        title = "{Optical Variability of Narrow-line and Broad-line Seyfert 1 Galaxies}",
      journal = {\apj},
         year = 2017,
        month = jun,
       volume = {842},
       number = {2},
          eid = {96},
        pages = {96},
          doi = {10.3847/1538-4357/aa72f4},
archivePrefix = {arXiv},
       eprint = {1705.05123},
 primaryClass = {astro-ph.GA},
       adsurl = {https://ui.adsabs.harvard.edu/abs/2017ApJ...842...96R}
}

@ARTICLE{1999ApJS..125..297L,
       author = {{Leighly}, Karen M.},
        title = "{A Comprehensive Spectral and Variability Study of Narrow-Line Seyfert 1 Galaxies Observed by ASCA. I. Observations and Time Series Analysis}",
      journal = {\apjs},
         year = 1999,
        month = dec,
       volume = {125},
       number = {2},
        pages = {297-316},
          doi = {10.1086/313277},
archivePrefix = {arXiv},
       eprint = {astro-ph/9907294},
 primaryClass = {astro-ph},
       adsurl = {https://ui.adsabs.harvard.edu/abs/1999ApJS..125..297L}
}

@ARTICLE{1992ApJS...80..109B,
       author = {{Boroson}, Todd A. and {Green}, Richard F.},
        title = "{The Emission-Line Properties of Low-Redshift Quasi-stellar Objects}",
      journal = {\apjs},
         year = 1992,
        month = may,
       volume = {80},
        pages = {109},
          doi = {10.1086/191661},
       adsurl = {https://ui.adsabs.harvard.edu/abs/1992ApJS...80..109B}
}

@ARTICLE{1982ApJ...255..419B,
       author = {{Blandford}, R.~D. and {McKee}, C.~F.},
        title = "{Reverberation mapping of the emission line regions of Seyfert galaxies and quasars.}",
      journal = {\apj},
         year = 1982,
        month = apr,
       volume = {255},
        pages = {419-439},
          doi = {10.1086/159843},
       adsurl = {https://ui.adsabs.harvard.edu/abs/1982ApJ...255..419B}
}

@ARTICLE{1993PASP..105..247P,
       author = {{Peterson}, Bradley M.},
        title = "{Reverberation Mapping of Active Galactic Nuclei}",
      journal = {\pasp},
         year = 1993,
        month = mar,
       volume = {105},
        pages = {247},
          doi = {10.1086/133140},
       adsurl = {https://ui.adsabs.harvard.edu/abs/1993PASP..105..247P}
}

@ARTICLE{2000ApJ...533..631K,
       author = {{Kaspi}, Shai and {Smith}, Paul S. and {Netzer}, Hagai and {Maoz}, Dan and {Jannuzi}, Buell T. and {Giveon}, Uriel},
        title = "{Reverberation Measurements for 17 Quasars and the Size-Mass-Luminosity Relations in Active Galactic Nuclei}",
      journal = {\apj},
         year = 2000,
        month = apr,
       volume = {533},
       number = {2},
        pages = {631-649},
          doi = {10.1086/308704},
archivePrefix = {arXiv},
       eprint = {astro-ph/9911476},
 primaryClass = {astro-ph},
       adsurl = {https://ui.adsabs.harvard.edu/abs/2000ApJ...533..631K}
}

@ARTICLE{2005ApJ...629...61K,
       author = {{Kaspi}, Shai and {Maoz}, Dan and {Netzer}, Hagai and {Peterson}, Bradley M. and {Vestergaard}, Marianne and {Jannuzi}, Buell T.},
        title = "{The Relationship between Luminosity and Broad-Line Region Size in Active Galactic Nuclei}",
      journal = {\apj},
         year = 2005,
        month = aug,
       volume = {629},
       number = {1},
        pages = {61-71},
          doi = {10.1086/431275},
archivePrefix = {arXiv},
       eprint = {astro-ph/0504484},
 primaryClass = {astro-ph},
       adsurl = {https://ui.adsabs.harvard.edu/abs/2005ApJ...629...61K}
}

@ARTICLE{2006ApJ...651..775B,
       author = {{Bentz}, Misty C. and {Denney}, Kelly D. and {Cackett}, Edward M. and {Dietrich}, Matthias and {Fogel}, Jeffrey K.~J. and {Ghosh}, Himel and {Horne}, Keith and {Kuehn}, Charles and {Minezaki}, Takeo and {Onken}, Christopher A. and {Peterson}, Bradley M. and {Pogge}, Richard W. and {Pronik}, Vladimir I. and {Richstone}, Douglas O. and {Sergeev}, Sergey G. and {Vestergaard}, Marianne and {Walker}, Matthew G. and {Yoshii}, Yuzuru},
        title = "{A Reverberation-based Mass for the Central Black Hole in NGC 4151}",
      journal = {\apj},
         year = 2006,
        month = nov,
       volume = {651},
       number = {2},
        pages = {775-781},
          doi = {10.1086/507417},
archivePrefix = {arXiv},
       eprint = {astro-ph/0607085},
 primaryClass = {astro-ph},
       adsurl = {https://ui.adsabs.harvard.edu/abs/2006ApJ...651..775B}
}

@ARTICLE{2009ApJ...697..160B,
       author = {{Bentz}, Misty C. and {Peterson}, Bradley M. and {Netzer}, Hagai and {Pogge}, Richard W. and {Vestergaard}, Marianne},
        title = "{The Radius-Luminosity Relationship for Active Galactic Nuclei: The Effect of Host-Galaxy Starlight on Luminosity Measurements. II. The Full Sample of Reverberation-Mapped AGNs}",
      journal = {\apj},
         year = 2009,
        month = may,
       volume = {697},
       number = {1},
        pages = {160-181},
          doi = {10.1088/0004-637X/697/1/160},
archivePrefix = {arXiv},
       eprint = {0812.2283},
 primaryClass = {astro-ph},
       adsurl = {https://ui.adsabs.harvard.edu/abs/2009ApJ...697..160B}
}

@ARTICLE{2013ApJ...767..149B,
       author = {{Bentz}, Misty C. and {Denney}, Kelly D. and {Grier}, Catherine J. and {Barth}, Aaron J. and {Peterson}, Bradley M. and {Vestergaard}, Marianne and {Bennert}, Vardha N. and {Canalizo}, Gabriela and {De Rosa}, Gisella and {Filippenko}, Alexei V. and {Gates}, Elinor L. and {Greene}, Jenny E. and {Li}, Weidong and {Malkan}, Matthew A. and {Pogge}, Richard W. and {Stern}, Daniel and {Treu}, Tommaso and {Woo}, Jong-Hak},
        title = "{The Low-luminosity End of the Radius-Luminosity Relationship for Active Galactic Nuclei}",
      journal = {\apj},
         year = 2013,
        month = apr,
       volume = {767},
       number = {2},
          eid = {149},
        pages = {149},
          doi = {10.1088/0004-637X/767/2/149},
archivePrefix = {arXiv},
       eprint = {1303.1742},
 primaryClass = {astro-ph.CO},
       adsurl = {https://ui.adsabs.harvard.edu/abs/2013ApJ...767..149B}
}

@software{2018ascl.soft09008G,
       author = {{Guo}, Hengxiao and {Shen}, Yue and {Wang}, Shu},
        title = "{PyQSOFit: Python code to fit the spectrum of quasars}",
 howpublished = {Astrophysics Source Code Library, record ascl:1809.008},
         year = 2018,
        month = sep,
          eid = {ascl:1809.008},
       adsurl = {https://ui.adsabs.harvard.edu/abs/2018ascl.soft09008G}
}

@ARTICLE{2020ApJ...891..178S,
       author = {{Sun}, Mouyuan and {Xue}, Yongquan and {Brandt}, W.~N. and {Gu}, Wei-Min and {Trump}, Jonathan R. and {Cai}, Zhenyi and {He}, Zhicheng and {Lin}, Da-bin and {Liu}, Tong and {Wang}, Junxian},
        title = "{Corona-heated Accretion-disk Reprocessing: A Physical Model to Decipher the Melody of AGN UV/Optical Twinkling}",
      journal = {\apj},
         year = 2020,
        month = mar,
       volume = {891},
       number = {2},
          eid = {178},
        pages = {178},
          doi = {10.3847/1538-4357/ab789e},
archivePrefix = {arXiv},
       eprint = {2002.08564},
 primaryClass = {astro-ph.HE},
       adsurl = {https://ui.adsabs.harvard.edu/abs/2020ApJ...891..178S}
}

@ARTICLE{2014Natur.513..210S,
       author = {{Shen}, Yue and {Ho}, Luis C.},
        title = "{The diversity of quasars unified by accretion and orientation}",
      journal = {\nat},
         year = 2014,
        month = sep,
       volume = {513},
       number = {7517},
        pages = {210-213},
          doi = {10.1038/nature13712},
archivePrefix = {arXiv},
       eprint = {1409.2887},
 primaryClass = {astro-ph.GA},
       adsurl = {https://ui.adsabs.harvard.edu/abs/2014Natur.513..210S}
}

@ARTICLE{2002ApJ...565...78B,
       author = {{Boroson}, Todd A.},
        title = "{Black Hole Mass and Eddington Ratio as Drivers for the Observable Properties of Radio-loud and Radio-quiet QSOs}",
      journal = {\apj},
         year = 2002,
        month = jan,
       volume = {565},
       number = {1},
        pages = {78-85},
          doi = {10.1086/324486},
archivePrefix = {arXiv},
       eprint = {astro-ph/0109317},
 primaryClass = {astro-ph},
       adsurl = {https://ui.adsabs.harvard.edu/abs/2002ApJ...565...78B}
}

@ARTICLE{2022MNRAS.514..164S,
       author = {{Stone}, Zachary and {Shen}, Yue and {Burke}, Colin J. and {Chen}, Yu-Ching and {Yang}, Qian and {Liu}, Xin and {Gruendl}, R.~A. and {Adam{\'o}w}, M. and {Andrade-Oliveira}, F. and {Annis}, J. and {Bacon}, D. and {Bertin}, E. and {Bocquet}, S. and {Brooks}, D. and {Burke}, D.~L. and {Carnero Rosell}, A. and {Carrasco Kind}, M. and {Carretero}, J. and {da Costa}, L.~N. and {Pereira}, M.~E.~S. and {De Vicente}, J. and {Desai}, S. and {Diehl}, H.~T. and {Doel}, P. and {Ferrero}, I. and {Friedel}, D.~N. and {Frieman}, J. and {Garc{\'\i}a-Bellido}, J. and {Gaztanaga}, E. and {Gruen}, D. and {Gutierrez}, G. and {Hinton}, S.~R. and {Hollowood}, D.~L. and {Honscheid}, K. and {James}, D.~J. and {Kuehn}, K. and {Kuropatkin}, N. and {Lidman}, C. and {Maia}, M.~A.~G. and {Menanteau}, F. and {Miquel}, R. and {Morgan}, R. and {Paz-Chinch{\'o}n}, F. and {Pieres}, A. and {Plazas Malag{\'o}n}, A.~A. and {Rodriguez-Monroy}, M. and {Sanchez}, E. and {Scarpine}, V. and {Serrano}, S. and {Sevilla-Noarbe}, I. and {Smith}, M. and {Suchyta}, E. and {Swanson}, M.~E.~C. and {Tarl{\'e}}, G. and {To}, C. and {DES Collaboration}},
        title = "{Optical variability of quasars with 20-yr photometric light curves}",
      journal = {\mnras},
         year = 2022,
        month = jul,
       volume = {514},
       number = {1},
        pages = {164-184},
          doi = {10.1093/mnras/stac1259},
archivePrefix = {arXiv},
       eprint = {2201.02762},
 primaryClass = {astro-ph.GA},
       adsurl = {https://ui.adsabs.harvard.edu/abs/2022MNRAS.514..164S}
}

@ARTICLE{2022ApJ...936..132Y,
       author = {{Yu}, Weixiang and {Richards}, Gordon T. and {Vogeley}, Michael S. and {Moreno}, Jackeline and {Graham}, Matthew J.},
        title = "{Examining AGN UV/Optical Variability beyond the Simple Damped Random Walk}",
      journal = {\apj},
         year = 2022,
        month = sep,
       volume = {936},
       number = {2},
          eid = {132},
        pages = {132},
          doi = {10.3847/1538-4357/ac8351},
archivePrefix = {arXiv},
       eprint = {2201.08943},
 primaryClass = {astro-ph.GA},
       adsurl = {https://ui.adsabs.harvard.edu/abs/2022ApJ...936..132Y}
}

@ARTICLE{1988ApJ...332..646A,
       author = {{Abramowicz}, M.~A. and {Czerny}, B. and {Lasota}, J.~P. and {Szuszkiewicz}, E.},
        title = "{Slim Accretion Disks}",
      journal = {\apj},
         year = 1988,
        month = sep,
       volume = {332},
        pages = {646},
          doi = {10.1086/166683},
       adsurl = {https://ui.adsabs.harvard.edu/abs/1988ApJ...332..646A}
}

@ARTICLE{2021ApJ...907...96S,
       author = {{Suberlak}, Krzysztof L. and {Ivezi{\'c}}, {\v{Z}}eljko and {MacLeod}, Chelsea},
        title = "{Improving Damped Random Walk Parameters for SDSS Stripe 82 Quasars with Pan-STARRS1}",
      journal = {\apj},
         year = 2021,
        month = feb,
       volume = {907},
       number = {2},
          eid = {96},
        pages = {96},
          doi = {10.3847/1538-4357/abc698},
archivePrefix = {arXiv},
       eprint = {2012.12907},
 primaryClass = {astro-ph.GA},
       adsurl = {https://ui.adsabs.harvard.edu/abs/2021ApJ...907...96S}
}

@ARTICLE{1999MNRAS.306..637G,
       author = {{Giveon}, Uriel and {Maoz}, Dan and {Kaspi}, Shai and {Netzer}, Hagai and {Smith}, Paul S.},
        title = "{Long-term optical variability properties of the Palomar-Green quasars}",
      journal = {\mnras},
         year = 1999,
        month = jul,
       volume = {306},
       number = {3},
        pages = {637-654},
          doi = {10.1046/j.1365-8711.1999.02556.x},
archivePrefix = {arXiv},
       eprint = {astro-ph/9902254},
 primaryClass = {astro-ph},
       adsurl = {https://ui.adsabs.harvard.edu/abs/1999MNRAS.306..637G}
}

@ARTICLE{2003A&A...398..927W,
       author = {{Wang}, J.-M. and {Netzer}, H.},
        title = "{Extreme slim accretion disks and narrow line Seyfert 1 galaxies: The nature of the soft X-ray hump}",
      journal = {\aap},
         year = 2003,
        month = feb,
       volume = {398},
        pages = {927-936},
          doi = {10.1051/0004-6361:20021511},
archivePrefix = {arXiv},
       eprint = {astro-ph/0210361},
 primaryClass = {astro-ph},
       adsurl = {https://ui.adsabs.harvard.edu/abs/2003A&A...398..927W}
}


\appendix
\renewcommand{\thefigure}{A\arabic{figure}}  
\setcounter{figure}{0}



\begin{figure*}
    \centering

    \includegraphics[width=0.48\linewidth]{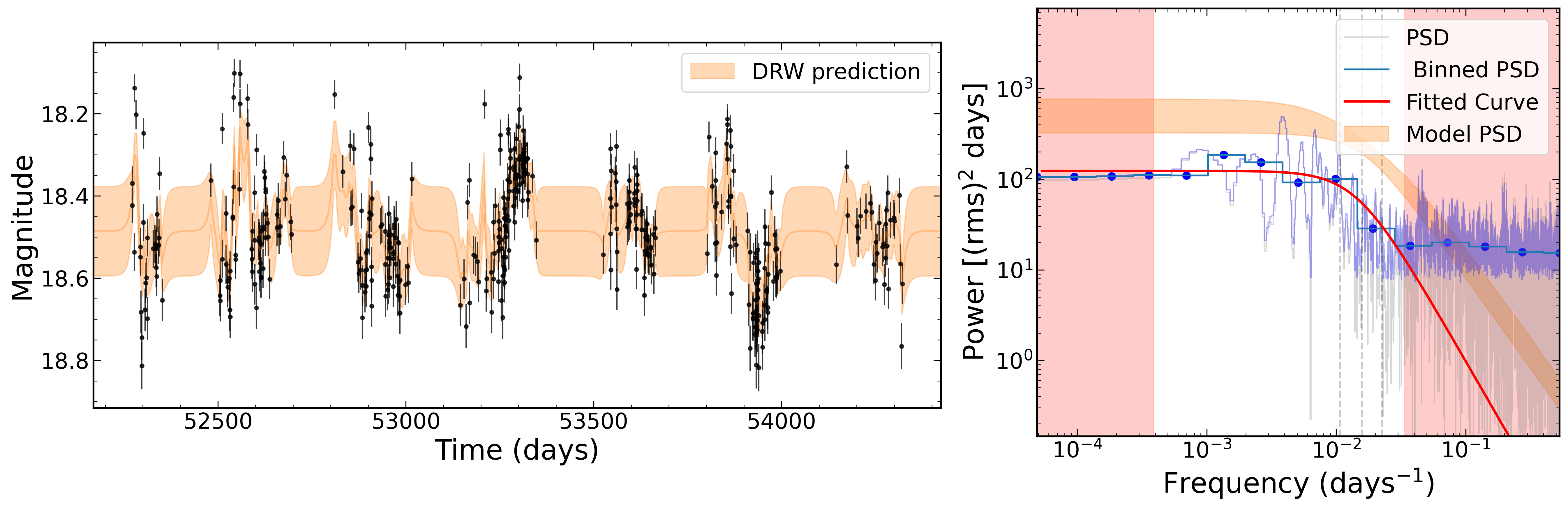}
    \hfill
    \includegraphics[width=0.48\linewidth]{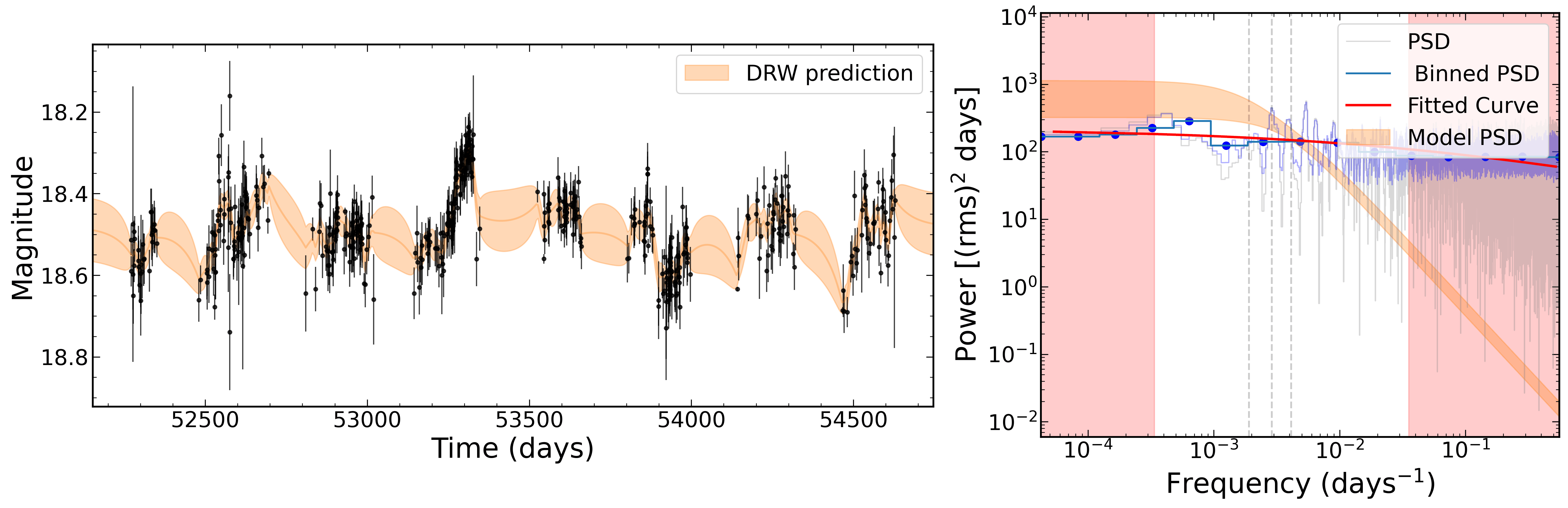}

    \vspace{1.5em}

    \includegraphics[width=0.48\linewidth]{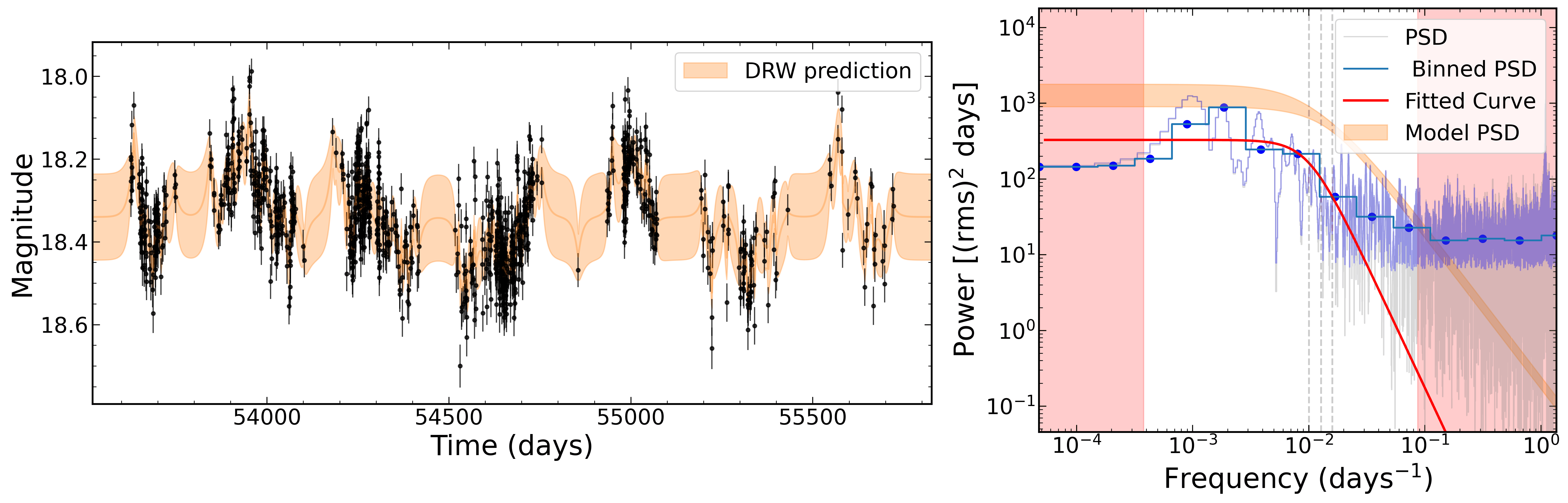}
    \hfill
    \includegraphics[width=0.48\linewidth]{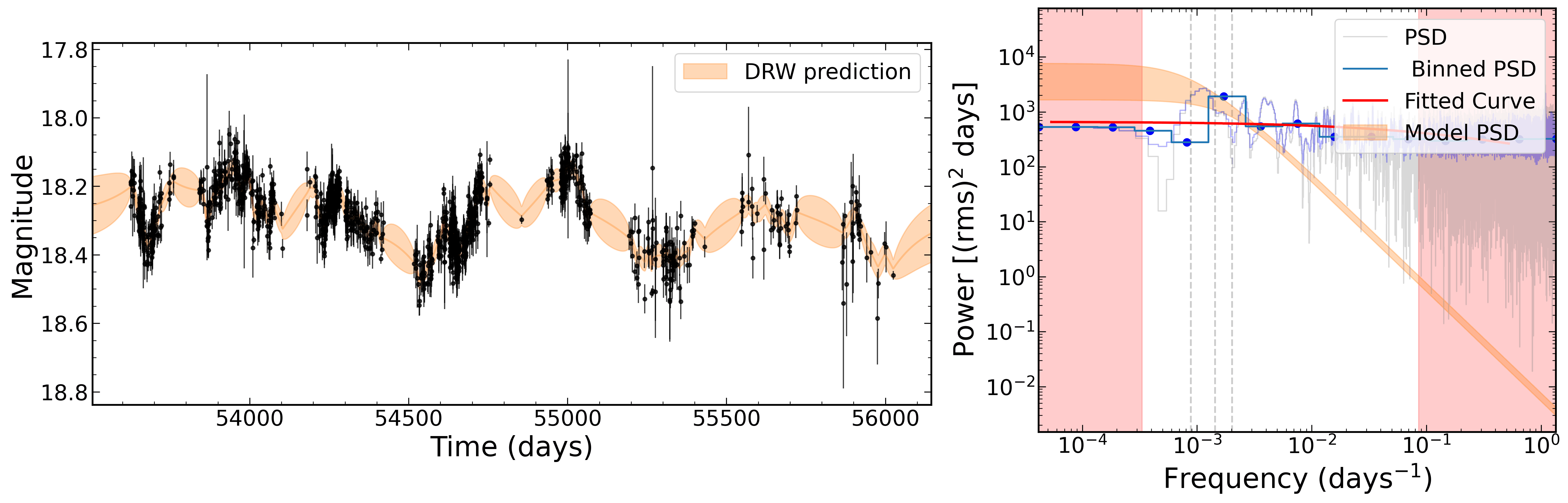}

    \vspace{1.5em}

    \includegraphics[width=0.48\linewidth]{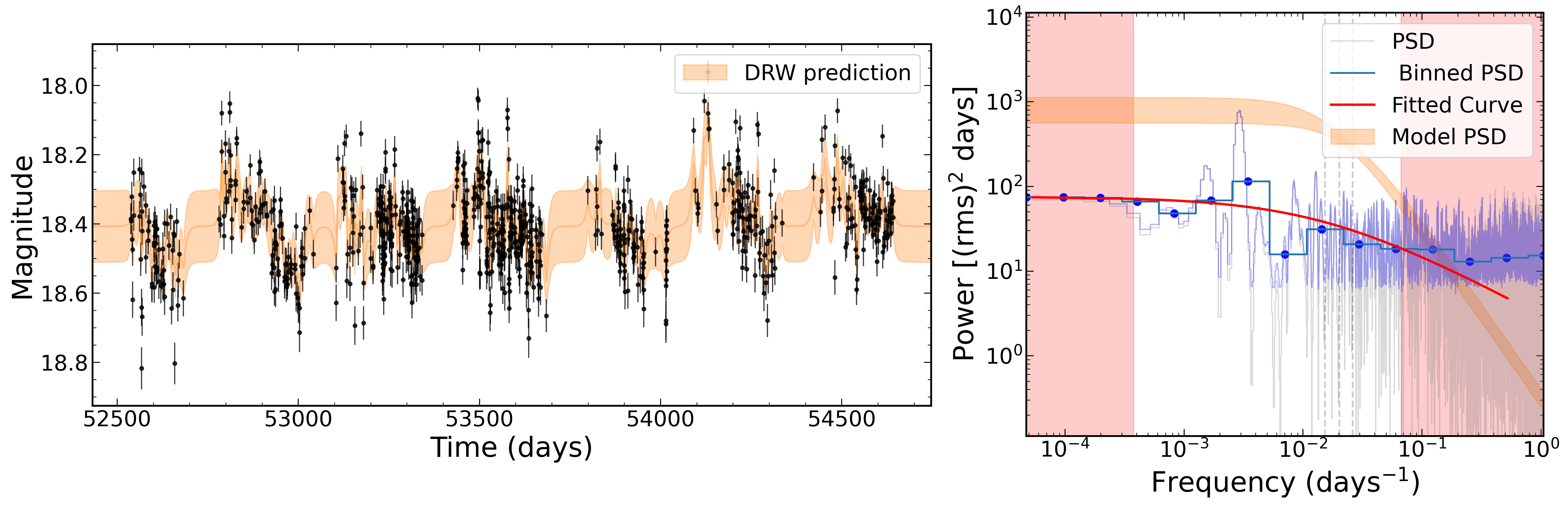}
    \hfill
    \includegraphics[width=0.48\linewidth]{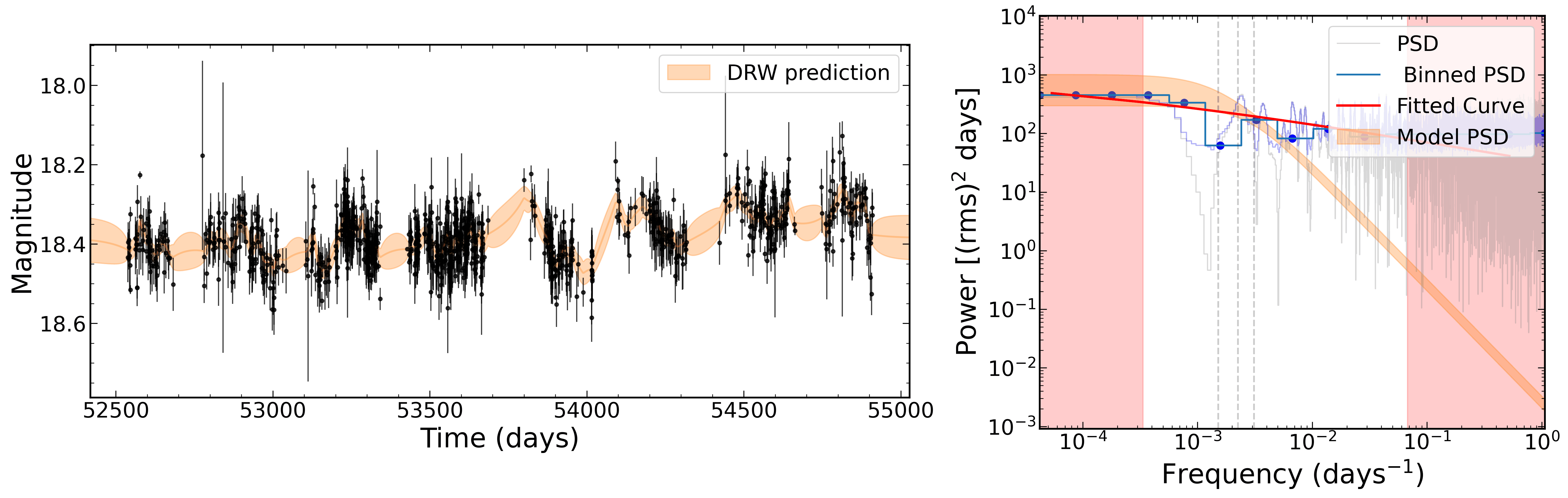}

    \vspace{1.5em}

    \includegraphics[width=0.48\linewidth]{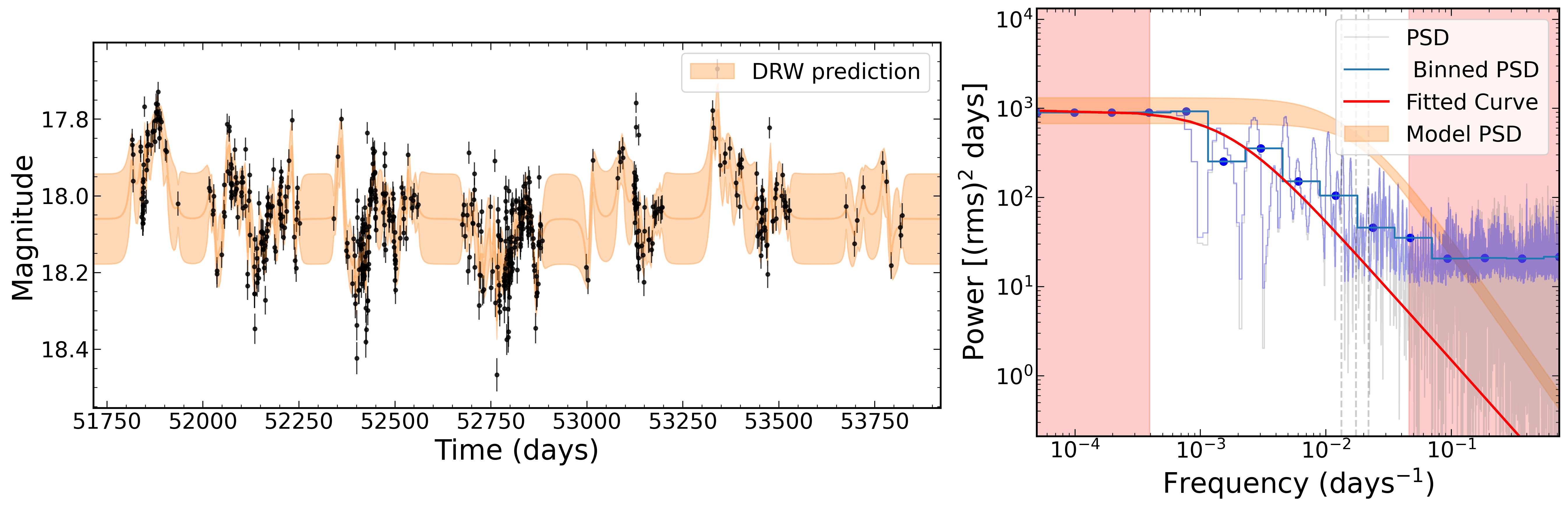}
    \hfill
    \includegraphics[width=0.48\linewidth]{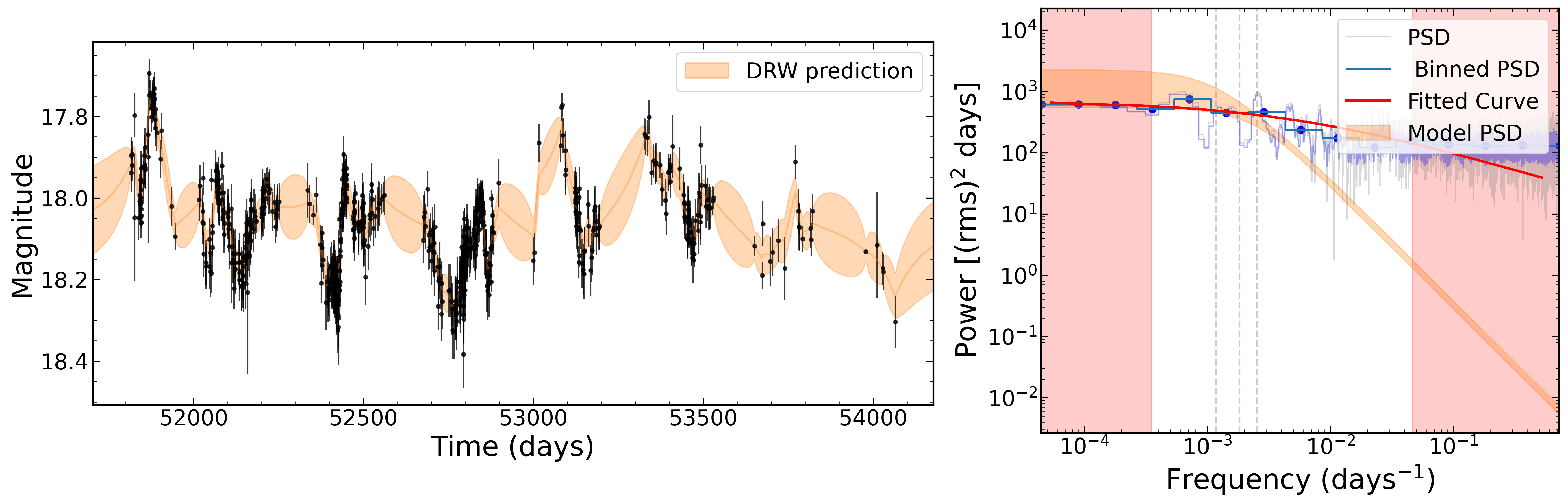}

    \caption{Comparison of ZTF $g$-band light curves between archival (normal) and forced photometry for four representative sources. Each row corresponds to a different object, identified by its right ascension (RA) and declination (Dec) coordinates: 
    (\textbf{Top to bottom}) 
    \texttt{(1) SDSSJ121948.93+054531.7 (RA,DEC = 184.9539, 5.758811)}, 
    \texttt{(2) SDSSJ130004.09+294624.4 (RA,DEC = 195.0171, 29.77346}, 
    \texttt{(3) SDSSJ144707.05+292025.5 (RA,DEC = 221.7794, 29.34043}, and 
    \texttt{(4) SDSSJ122900.29+272522.2 (RA,DEC = 187.2512, 27.42285}. 
    In each pair, the \textit{left panel} shows the light curve and corresponding DRW+PSD fit based on archival photometry, while the \textit{right panel} shows the same analysis using forced photometry. These comparisons highlight the improved sampling and PSD fidelity achieved with forced photometry, enhancing the reliability of DRW-based modeling.}

    \label{fig:photometry_comparison_all}
\end{figure*}

\addtocounter{figure}{0}
\begin{figure}
    \centering
    \includegraphics[width=1\linewidth]{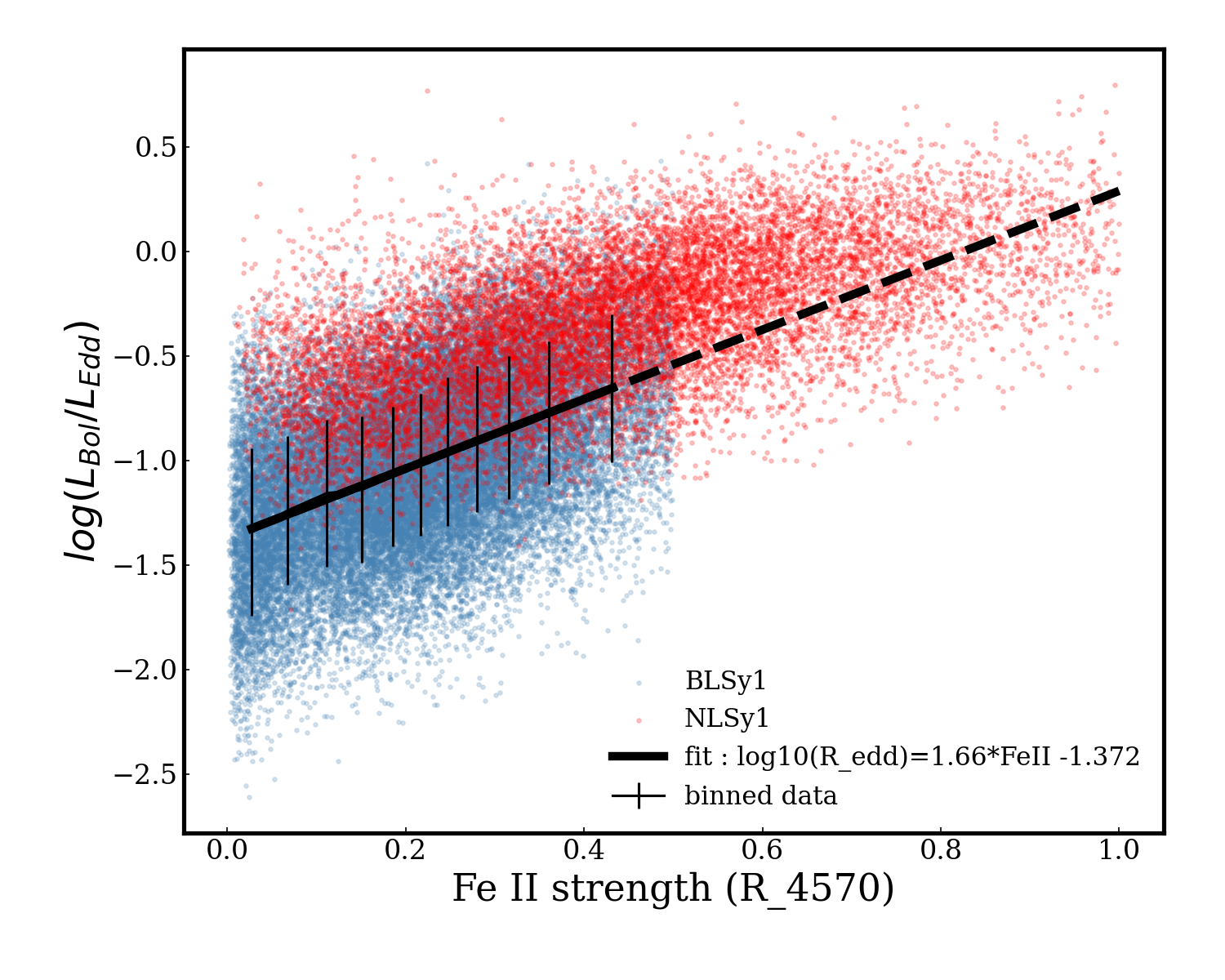}
    \caption{Scatter plot of Eddington ratio and FeII strength (R4570) for the full sample of NLSy1 (blue) and BLSy1 (red) galaxies. The solid black line denotes the best-fit linear relation for the BLSy1 subsample, extended to the NLSy1 regime as a dashed line. The best-fit relation is $\log(\text{L}_{\text{Bol}} / \text{L}_{\text{Edd}}) = 1.66 \times \text{R}_{4570} - 1.372$.}
    \label{fig:FeII_strength}
\end{figure}

\clearpage
\section*{Table of First 10 Sources}

\begin{sidewaystable}[htbp]
    \centering
    \small
    \resizebox{\textwidth}{!}{\renewcommand{\arraystretch}{1.3}
\setlength{\tabcolsep}{3pt}
\scriptsize
\begin{tabular}{|c | c | c | c | c | c | c | c | c | c | c | c | c | c | c | c | c | c|}
\hline
SDSS Name & RA & DEC & \makecell{Redshift\\ (z)} & \makecell{Median\\ Error (mag)} & \makecell{$ \log_{10} \; M_{\mathrm{BH, cele}} $\\( [M$_\odot$])} & \makecell{$ \log_{10} \; M_{\mathrm{BH, SE}} \pm \Delta M_{\mathrm{BH}}$\\( [M$_\odot$])} & \makecell{$\tau_{\mathrm{rec,cele}}$\\(days)} & $\sigma$ (mag) & \makecell{Zitter\\(mag)} & p-value & \makecell{Bad Region \\(days)} & \makecell{$ \log_{10} \;L_{\mathrm{bol}} \pm \Delta L_{\mathrm{bol}}$\\([erg/s])} & $\log_{10} \; R_{\mathrm{Edd}} \pm \Delta R_{\mathrm{Edd}}$ & $R4570$ \\ \hline
084046.96+400647.0 & 130.1957 & 40.1131 & 0.23 & 0.05 & 8.08 & 6.61 ± 0.03 & 110.65 & 0.07 & 0.00 & 0.00 & 389.94 & 44.87 ± 0.02 & 0.12 ± 0.03 & 0.76 \\
152447.13+520759.1 & 231.1964 & 52.1331 & 0.16 & 0.03 & 7.69 & 6.70 ± 0.08 & 77.53 & 0.08 & 0.03 & 0.00 & 453.15 & 44.00 ± 0.17 & -0.84 ± 0.19 & 0.34 \\
022608.20-005318.8 & 36.5342 & -0.8886 & 0.11 & 0.03 & 8.23 & 6.72 ± 0.05 & 127.08 & 0.08 & 0.00 & 0.00 & 413.19 & 44.23 ± 0.05 & -0.62 ± 0.07 & 0.11 \\
163142.49+465243.2 & 247.9271 & 46.8787 & 0.19 & 0.07 & 8.08 & 6.72 ± 0.08 & 110.29 & 0.11 & 0.00 & 0.00 & 403.66 & 44.04 ± 0.14 & -0.82 ± 0.17 & 0.37 \\
123556.54+015728.2 & 188.9856 & 1.9578 & 0.25 & 0.06 & 7.61 & 6.61 ± 0.14 & 72.37 & 0.09 & 0.03 & 0.00 & 421.08 & 44.25 ± 0.19 & -0.50 ± 0.23 & 0.75 \\
101832.19+623959.8 & 154.6341 & 62.6666 & 0.35 & 0.05 & 8.13 & 6.74 ± 0.03 & 115.76 & 0.06 & 0.00 & 0.00 & 355.40 & 45.07 ± 0.04 & 0.19 ± 0.05 & 0.47 \\
133059.08+602128.3 & 202.7462 & 60.3579 & 0.29 & 0.05 & 7.96 & 6.60 ± 0.14 & 98.98 & 0.09 & 0.00 & 0.00 & 359.97 & 44.22 ± 0.30 & -0.52 ± 0.34 & 0.25 \\
083949.64+484701.4 & 129.9569 & 48.7837 & 0.04 & 0.03 & 8.09 & 6.18 ± 0.05 & 111.81 & 0.07 & 0.00 & 0.00 & 503.46 & 43.53 ± 0.11 & -0.79 ± 0.12 & 0.35 \\
215101.17-083716.2 & 327.7549 & -8.6212 & 0.12 & 0.02 & 8.10 & 6.83 ± 0.01 & 112.27 & 0.08 & 0.01 & 0.00 & 417.39 & 44.80 ± 0.01 & -0.17 ± 0.02 & 0.52 \\
090628.20+455841.5 & 136.6175 & 45.9782 & 0.18 & 0.06 & 7.60 & 6.29 ± 0.07 & 71.22 & 0.09 & 0.00 & 0.00 & 406.02 & 44.17 ± 0.13 & -0.27 ± 0.14 & 0.68 \\
\hline
\end{tabular}}
    \caption{Tabulated properties for the first 10 NLSy1 sources in the sample.}
    \label{tab:table}
\end{sidewaystable}

\bsp	
\label{lastpage}
\end{document}